\documentclass[12pt,preprint]{aastex}

\usepackage{lscape}

\newcommand{\etal}{{et al.}}

\shorttitle{A new method for Astrometric mm-VLBI}
\shortauthors{Rioja \& Dodson}

\begin{document}

\title{High Precision Astrometric Millimeter VLBI Using a New Method for
Atmospheric Calibration.}

\author{M. Rioja\altaffilmark{1,2} and R. Dodson\altaffilmark{1}}
\affil{$^1$ ICRAR, UWA, Perth, Australia}
\email{maria.rioja@icrar.org}

\altaffiltext{2}{On secondment Observatorio Astron\'omico Nacional
  (OAN), Spain.} 

\begin{abstract}

  We describe a new method which achieves high precision Very Long
  Baseline Interferometry (VLBI) astrometry in observations at
  millimeter wavelengths.  It combines fast frequency-switching
  observations, to correct for the dominant non-dispersive
  tropospheric fluctuations, with slow source-switching observations,
  for the remaining ionospheric dispersive terms. We call this method
  Source-Frequency Phase Referencing.  Provided that the switching
  cycles match the properties of the propagation media, one can
  recover the source astrometry.

  We present an analytic description of the two-step calibration
  strategy, along with an error analysis to characterize its
  performance.  Also, we provide observational demonstrations of a
  successful application with observations using the Very Long Baseline
  Array at 86 GHz of the pairs of sources 3C274 \& 3C273 and
  1308+326 \& 1308+328, under various conditions.

  We conclude that this method is widely applicable to millimeter VLBI
  observations of many target sources, and unique in providing {\it
    bona-fide} astrometrically registered images and high precision
  relative astrometric measurements in mm-VLBI using existing and
  newly built instruments.
\end{abstract}


\section{Introduction}\label{sec:intro}

The comparative study of the radiation emitted at multiple radio bands
has proved to be a useful tool in astronomy for the investigation of
the nature of the emission mechanisms and to probe the physical
conditions of the emitting regions.  Multi-frequency observations with
the high spatial resolution obtained with Very Long Baseline
Interferometry (VLBI) are suitable for the study of extragalactic
radio sources, such as AGNs, providing detailed images of the
radiation from the relativistic jets, which are launched from the
central engine that powers the sources.  Observations at increasingly
high frequencies offer the prospect of an increasingly deep
exploration of the inner jet region, closer to the central engine.

By comparing well aligned high resolution images at multiple
frequencies it is possible to map the spectral index across the jet
structure. The spectral index map carries direct information about the
physical conditions in the jet regions, and, potentially, with
observations at the highest frequencies, on the structure of the
central engine \citep{bh_shadow_1}.  Also, the standard model
\citep{blandford_79} predicts changes in the apparent position of the
observed ``core'' component, at the base of the jet, in observations
at different frequencies as a result of opacity effects in the jet.
These position changes are called {\it core-shifts} and hold a direct
relationship with the conditions in the nuclear region at the base of
the jet where the ``core'' is located.  For both studies the precise
alignment of the source images is mandatory to assess true intrinsic
source properties using multi-frequency comparison techniques,
otherwise alignment errors will result in misleading conclusions.

Standard VLBI images, which are created using self-calibration
techniques, provide exquisite detail on the source structure but lack
astrometric information.  The astrometry is lost in the process of
removing the residual contributions arising from imprecise modeling
of the propagation effects through the atmosphere and the use of
independent frequency standards at each telescope, among others.

The special analysis technique of Phase Referencing ({\em hereafter}
PR) is required to preserve the astrometric
information. PR relies in the use of interleaving observations of an
external calibrator source to correct the errors present in the target
dataset, rather than using the target data themselves as in standard
VLBI analysis \citep{alef88}. 
By doing this it is possible to achieve high-precision
(relative) {\it bona-fide} astrometric measurements of the angular
separation between the two sources.  The switching time and switching
angle are critical parameters for the success of PR techniques.
Typical switching values are estimated using the temporal and spatial
structure-function of the atmospheric fluctuations, and are dependent
on the observing frequency \citep{memo_20}.  Conventional phase
referencing has been successfully used at cm-wavelengths (from 1.4 to
43-GHz) for which atmospheric effects are moderate and calibrator
sources are easy to find.

VLBI observations at mm-wavelengths are challenging because of the
lower sensitivity of the instruments, intrinsically lower source
fluxes and shorter coherence times imposed by the rapid variations of
the water vapor content in the troposphere.  For the same reasons,
VLBI astrometry with conventional PR at high frequencies, beyond 43
GHz, is practically impossible due to the extremely short telescope
switching times involved. The only successful demonstration was with
the VLBA at 86-GHz for a pair of sources only 14$^\prime$ apart
\citep{porcas_02_pr86,porcas_03_pr86}.

An alternative approach to overcome the tropospheric limitations in
observations at high frequencies consists in using fast frequency-switching, instead of fast source-switching as in PR, on the grounds
that the tropospheric excess path delay, being independent of the
observing frequency (i.e. it is a non-dispersive medium), can be
corrected for using dual frequency observations.  This technique has
been attempted in VLBI resulting in longer effective coherence
times at mm-wavelengths, but it failed to recover astrometry due to
remaining dispersive ionospheric and instrumental errors
\citep{middelberg_05_fs}.

We propose that astrometry at high frequencies can be achieved by
combining alternating observations at two frequencies, to correct for
the non-dispersive propagation media effects, and of two sources, to
correct for the remaining dispersive effects, providing suitable
switching times and switching angles are used.  We term this new
technique {\it Source-Frequency Phase Referencing} ({\em hereafter}
SFPR).  The direct outcomes of this technique are: high precision
 {\it bona-fide} astrometric measurements of the angular
separation between emitting regions in the two frequency bands, and
increased coherence time in VLBI observations at the highest
frequencies.  Hence it allows {\it bona-fide} astrometric registration
of VLBI maps in the high frequency regime, beyond the threshold for
conventional PR techniques.  For example, applied to AGN-jets it would allow
spectral index and core-shift measurements; applied to spectral line
VLBI observations this would allow the alignment of the spatial
distribution of emission arising from multiple maser transitions of a
given molecule.  Such information is of great interest in astrophysics
(e.g.  \cite{lob_98_cj,m87_dodson,soria_07,rioja_pasj}). Moreover, the
combination of SFPR and conventional PR techniques holds the prospect
of providing high precision relative astrometric measurements of 
positions with respect to an external reference (i.e. a
calibrator source).  This would enable VLBI multi-epoch proper motion
and parallax studies at the highest frequencies.

This paper presents an analytical description of this new technique
that enables high precision VLBI astrometric measurements in the
highest frequency regime, {along with an experimental demonstration
  using VLBA observations at 43 and 86\,GHz}.  Also, we present a
comparative error analysis and discuss the feasibility of the new
technique in the context of existing and newly built instruments.  A
comprehensive computer-simulation study of the SFPR performance will
be presented elsewhere.

\section{The Method}\label{sec:method}

\begin{figure}
\epsscale{1}
\plotone{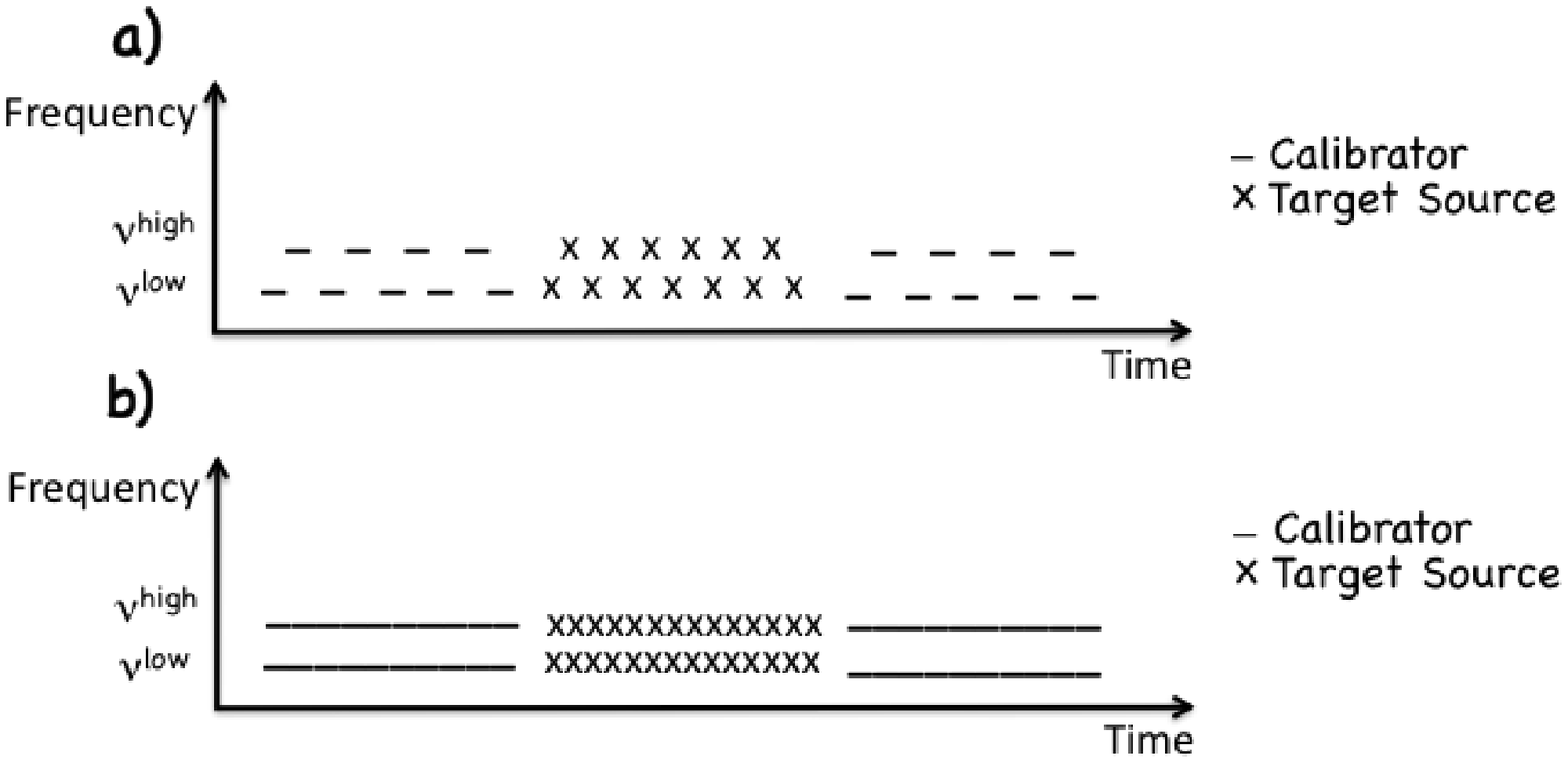}
\caption{A sketch showing the distribution of observing time
  per frequency and per source  (typically minutes), in SFPR observations
  using: a) fast frequency-switching with slow source-switching, and
  b) simultaneous dual-frequency observations with slow
  source-switching. }
\label{fig1}
\end{figure}

This section contains an analytic description of the basis of the SFPR
astrometric technique. SFPR uses interleaving observations at a
different frequency, and of a different source, to compensate for
non-dispersive and dispersive errors, respectively, in the target
observations.  This approach resembles conventional PR
techniques in the use of external observations to derive the
calibration, rather than the target data themselves.  SFPR can be applied
to a wide range of frequencies, providing the switching cycles and
switching angle match the fluctuations of the residual errors in the
analysis.  The frequency switching cycle ($T^{\nu}_{swt}$) corresponds to the
elapsed time between midpoints of two consecutive scans at the same frequency, and
the source switching cycle ($T_{swt}$) is between blocks of scans on the same
source.  In the ideal case when multiple frequencies can be observed
simultaneously ($T^{\nu}_{swt}$=0) only source switching is
required. The
switching angle is the angular separation between the two sources.
Figure \ref{fig1} describes the allocation of observing time, for each of the
frequencies and sources, in SFPR observations.  In this paper we will
focus on the high frequency regime, where the dominant rapid
tropospheric (non-dispersive) fluctuations prevent the application of
conventional PR techniques, and SPFR is unique in its astrometric
application.  As a general rule, a fast frequency-switching cycle
combined with a slow source-switching cycle is appropriate for SFPR
observations at high frequencies.  Section \ref{sec:guide} contains detailed
guidelines for planning such SFPR observations.

In order to simplify the presentation, we will use hereafter 
the term ``target'' to refer to both the source and frequency of
interest; ``reference frequency'', or simply ``reference'', for the
other frequency; ``calibrator source'', or simply ``calibrator'', for
the other source. In the formulae, we will use the superscripts $high$
and $low$ to refer to the target and reference frequencies, 
respectively; and subscripts $A$ and $B$ for the target and
calibrator sources, respectively.

The SFPR calibration strategy comprises of two steps.  The first step
(subsection \ref{sec:m_step1}) assumes that the dominant tropospheric residual
errors (and in general, any non-dispersive errors) in the target
dataset can be removed using the observations at a lower
reference frequency, on the same source. The second step (subsection
\ref{sec:m_step2}) assumes that the remaining ionospheric and instrumental errors
(and in general, any other dispersive errors) can be removed using the
interleaved observations of an external calibrator source.  Finally,
the nature of the astrometry enabled with this combined calibration
technique is described in subsection 2.3.

\subsection{\it {{\it Step 1:} Calibration of  Tropospheric
 errors using fast frequency-switching on the same
source}}\label{sec:m_step1}

The first SFPR calibration step uses fast-frequency switching 
observations, alternating between a lower reference 
frequency ($\nu^{low}$) and a higher target frequency ($\nu^{high}$),
to compensate for the effect of errors in the tropospheric delay model 
at the target frequency.
Such errors introduce an excess delay which is
independent of the observing frequency (i.e. non-dispersive).

Our description of the SFPR method uses the residual phase VLBI
observable, i.e.  after the {\it a\,priori} model values for the
various contributing terms have been removed at the correlation and the
signal has been integrated, for each baseline.  Following standard
nomenclature the residual phase error values for observations of the
target source ($A$) at the reference frequency, $\phi^{low}_{A}$, are
shown as a sum of contributions:

\begin{equation}
  \phi^{{\rm low}}_{{\rm A}} = \phi^{{\rm low}}_{{\rm A,str}} + \phi^{{\rm low}}_{{\rm A,geo}} + 
  \phi^{{\rm low}}_{{\rm A,tro}} +\phi^{{\rm low}}_{{\rm A,ion}}
  +\phi^{{\rm low}}_{{\rm A,inst}} + \phi^{{\rm low}}_{{\rm A,thermal}} + 
  2\pi n^{{\rm low}}_{{\rm A}}, \hspace*{1cm}
  n^{{\rm low}}_{\rm A} \in {\rm integer}, 
\end{equation}

\noindent
where $\phi^{{\rm low}}_{{\rm A,geo}}, \phi^{{\rm low}}_{{\rm A,tro}}, \phi^{{\rm low}}_{{\rm A,ion}}$,
and $\phi^{{\rm low}}_{{\rm A,inst}}$ are the contributions arising from
geometric, tropospheric, ionospheric and instrumental inadequacies in
the delay model, respectively.  The $\phi^{{\rm low}}_{{\rm A,str}}$ term is the
visibility phase, which accounts for the contribution of the source
structure to the observed phases.  Point-like source structures, and
more generally symmetric structures, have $\phi^{{\rm low}}_{{\rm A,str}}=0$.  The
$\phi^{{\rm low}}_{{\rm A,thermal}}$ term is the measurement error due to the
sensitivity of the array, and is usually much smaller than the rest of
the contributions.  The $2\pi n^{{\rm low}}_{{\rm A}}$ term, with $n^{{\rm low}}_{{\rm A}}$ an
integer value, stands for the inherent modulo $2 \pi$ phase ambiguity.

Similarly, for the target dataset, the residual phase error values for
the interleaving observations of the target source at the 
target frequency, $\phi^{{\rm high}}_{{\rm A}}$, are:

\begin{displaymath}
  \phi^{{\rm high}}_{{\rm A}} = \phi^{{\rm high}}_{{\rm A,str}} + \phi^{{\rm high}}_{{\rm A,geo}} + 
  \phi^{{\rm high}}_{{\rm A,tro}} +\phi^{{\rm high}}_{{\rm A,ion}}
  +\phi^{{\rm high}}_{{\rm A,inst}} +  \phi^{{\rm high}}_{{\rm A,thermal}} +  2\pi n^{{\rm high}}_{{\rm A}},
\end{displaymath}

with contributions as described above for Eq. 1, at the 
target frequency, $\nu^{{\rm high}}$. \\
The tropospheric phase errors are proportional to the observing
frequency, as for all non-dispersive media, and 
are the dominant contribution at the 
high frequencies. Instead, the dispersive ionospheric 
phase error is inversely proportional to the observing
frequency, hence its effect becomes weaker as frequency increases.

The SFPR analysis starts by applying standard VLBI self-calibration
and hybrid imaging techniques to the observations of the target source
(${\rm A}$) at the reference frequency ($\nu^{{\rm low}}$).  This produces a map
of the source along with a set of antenna-based terms,
$\phi^{{\rm low}}_{{\rm A,self-cal}}$, that account for the sum of the errors in
Eq. 1, except for $\phi^{{\rm low}}_{{\rm A,str}}$.

Next, these antenna-based corrections are interpolated to the
interleaving scan times when the target frequency is observed, $\tilde
\phi^{{\rm low}}_{{\rm A,self-cal}}$, and scaled by the frequency ratio $R$, with
$R = {{\rm \nu^{high}} \over \nu^{{\rm low}}}$.  The resultant values,
$R\,.\,\tilde \phi^{{\rm low}}_{{\rm A,self-cal}}$, provide the basis for the
tropospheric calibration of the target dataset.
We call the resultant calibrated target dataset, using the
interpolated plus scaled estimated solutions from the observations at
the reference frequency, ``Frequency Phase Transferred'' ({\it
  hereafter} FPT), or {\it troposphere-free} target dataset. The
corresponding residual target phases, $\phi_{\rm A}^{{\rm FPT}}$, are:

\begin{center}
\begin{eqnarray}
\phi_{\rm A}^{{\rm FPT}} = 
\phi_{\rm A}^{{\rm high}} - R\,.\,\tilde \phi_{{\rm A,self-cal}}^{{\rm low}} 
= \phi_{{\rm A,str}}^{{\rm high}} \nonumber \\ 
+ (\phi_{{\rm A,geo}}^{{\rm high}} - R\,.\,\tilde \phi_{{\rm A,geo}}^{{\rm low}}) 
+ (\phi_{{\rm A,tro}}^{{\rm high}} - R\,.\,\tilde \phi_{{\rm A,tro}}^{{\rm low}}) \nonumber \\
+ (\phi_{{\rm A,ion}}^{{\rm high}} - R\,.\,\tilde \phi_{{\rm A,ion}}^{{\rm low}})  
+ (\phi_{{\rm A,inst}}^{{\rm high}} - R\,.\,\tilde \phi_{{\rm A,inst}}^{{\rm low}}) \nonumber \\
+ 2\pi (n^{{\rm high}}_{\rm A}-R\,.\,n^{{\rm low}}_{\rm A})  ,
\end{eqnarray} 
\end{center}

where we have omitted the thermal noise term, for simplicity.

At this point we can make some reasonable approximations concerning
the difference terms in brackets in Eq. 2. 
First, given that the tropospheric residual phase errors scale
linearly with frequency, and providing that the frequency switching
cycle used is suitable to sample the rapid tropospheric fluctuations
then:

\begin{displaymath}
\phi_{{\rm A,tro}}^{{\rm high}} - R\,.\,\tilde \phi_{{\rm A,tro}}^{{\rm low}} \approx 0 ,
\end{displaymath}

Second, given that the geometric errors are also non-dispersive, the
effect of any errors in the antenna and source coordinates will
effectively cancel out using this calibration procedure, except for a
frequency-dependent source position shift, such as core-shift
phenomena in AGNs, but in general any change in position between the
frequency bands irrespective of its nature. We refer to all of these
as ``core-shifts'' hereafter:

\begin{eqnarray} 
\phi^{{\rm high}}_{{\rm A,geo}}  - R\,.\,\tilde\phi^{{\rm low}}_{{\rm A,geo}}  \approx
2\pi \,\vec{D}_{\lambda^{{\rm high}}}\,.\,\vec{\theta}_{{\rm A}} \nonumber , 
\end{eqnarray} 

\noindent
where $\vec{\theta}_{{\rm A}}$ stands for the target source ``core-shift'' 
between the two observed frequencies, 
and $\vec{D}_{\lambda^{{\rm high}}}$ is the interferometer 
baseline vector in units of 
$\lambda^{{\rm high}}$ wavelengths, with $\lambda^{{\rm high}} = c / \nu^{{\rm high}}$.

However, the dispersive residual errors in the target dataset 
will not be compensated using this calibration procedure.
The ionospheric and instrumental residual phase errors
belong to this category, and 
therefore remaining residuals are expected:

\begin{center}
\begin{equation}
\phi_{{\rm A,ion}}^{{\rm high}} - R\,.\,\tilde \phi_{{\rm A,ion}}^{{\rm low}} \approx
({1 \over R}-R)\, \tilde \phi_{{\rm A,ion}}^{{\rm low}} 
\eqnum{3}
\end{equation}
\end{center}

\begin{center}
\begin{displaymath}
\phi_{{\rm A,inst}}^{{\rm high}} - R\,.\,\tilde \phi_{{\rm A,inst}}^{{\rm low}} \neq 0 .
\end{displaymath}
\end{center}

\noindent
Making use of the approximations described above, Eq. 2 becomes:

\begin{equation}
\phi_{\rm A}^{{\rm FPT}} = 
 \phi_{{\rm A,str}}^{{\rm high}} + 
2 \pi \, \vec{D}_{\lambda^{{\rm high}}}\,.\, \vec{\theta}_{{\rm A}}
+ ({1 \over R}-R) \, \tilde \phi_{{\rm A,ion}}^{{\rm low}} 
+ (\phi_{{\rm A,inst}}^{{\rm high}} - R \,.\, \tilde \phi_{{\rm A,inst}}^{{\rm low}})
+ \Delta_{{\rm i,T^\nu_{\rm swt}}}, \,\,\,\,\,\,\, R \in {\rm integer} ,
\eqnum{4}
\end{equation}

\noindent
where $\Delta_{{\rm i,T^\nu_{\rm swt}}}$ stands for the interpolation errors
arising from using a frequency switching cycle $T^\nu_{{\rm swt}}$. Note
that with simultaneous dual frequency observations, for example using
the Korean VLBI Network (KVN) \citep{kvn_ref}, no interpolation is required and
$\Delta_{{\rm i,T^\nu_{swt}}}$=0. For simplicity we have omitted the $2\pi$
phase ambiguity term in Eq. 4 which, providing $R$ is an integer
value, just adds an unknown number of complete turns and is irrelevant
for the analysis.  The importance of having an integer ratio between
the frequencies involved in SFPR calibration is discussed in Section
\ref{sec:guide}.

Eq. 4 shows that the {\sc fpt}-calibrated target dataset is free of
the random and systematic tropospheric errors, providing a suitably
fast frequency-switching cycle was used at the observations.  However,
contaminating long-term ionospheric and instrumental residual phase
variations still remain blended with the source structure and the
astrometric ``core-shift'' signature, preventing its direct
extraction.
At this point, previous realizations of the dual-frequency VLBI
calibration method applied self-calibration techniques on the {\sc
  fpt}-calibrated target dataset to eliminate the contaminating terms,
with the consequent loss of the astrometric information
\citep{middelberg_05_fs, jung_evn}.

We propose instead an alternative calibration strategy which preserves
the chromatic astrometry signature (i.e. the ``core-shift''
information). This is described in the next section.

\subsection{\it {{\it Step 2}: Calibration of Ionospheric errors
using slow telescope-switching between two sources.}}\label{sec:m_step2}

The second SFPR calibration step uses the observations of an external
calibrator source ({\it B}) to eliminate the remaining dispersive
errors in Eq. 4, in a similar fashion as in conventional phase
referencing.  The observations of the calibrator and
the target sources are interleaved using a much slower switching cycle
than that required for conventional phase referencing. A discussion
on the constraints on the source switching cycle and the switching
angle is presented in Section \ref{sec:guide}.
The observations of the calibrator source are carried out using
fast-frequency switching scans, as described for the target source,
and analyzed following the same procedures
as described in section \ref{sec:m_step1} for the target source.
Therefore, the tropospheric-free residual phases for the 
calibrator source, $\phi_{{\rm B}}^{{\rm FPT}}$, can be expressed, following Eq. 4, as: 

\begin{equation}
\phi_{{\rm B}}^{{\rm FPT}} = 
2 \pi \, \vec{D}_{\lambda^{{\rm high}}}\,.\,\vec{\theta}_{{\rm B}}
+ ({1 \over R}-R) \, \tilde \phi_{{\rm B,ion}}^{{\rm low}}
+ (\phi_{{\rm B,inst}}^{{\rm high}} - R \,.\, \tilde \phi_{{\rm B,inst}}^{{\rm low}})
+ \Delta_{{\rm i,T^{\nu}}_{\rm swt}} ,
\eqnum{5}
\end{equation}

\noindent
where $\vec{\theta}_{{\rm B}}$ is the ``core-shift'' in the calibrator, and
the rest of the terms are similar to those described in Eq. 4, but for the
calibrator source. 
We assume a compact calibrator source, 
i.e. $\phi_{{\rm B,str}}^{{\rm high}}=0$; in general one should correct for the 
structure term, $\phi_{{\rm B,str}}^{{\rm high}}$, using the calibrator hybrid map.

It is reasonable to assume that the tropospheric-free calibrator
dataset (Eq. 5) could be used to remove the remaining errors in the
tropospheric-free target dataset (Eq. 4), as done in conventional PR
analysis.  This is the essence of the {\it Step 2} calibration.  We
dubbed the resultant differenced values the ``{\sc sfpr}-calibrated''
dataset.

The conditions under which the {\it Step 2} calibration 
works are:

{\it a)} that the source switching cycle is faster than the residual
ionospheric fluctuations at the reference frequency, and {\it b)} that
the angular separation between the calibrator and target sources is
smaller than the ionospheric isoplanatic patch-size at the reference
frequency, which is defined as the area over which the variation of
excess phase due to the ionosphere is small compared with $2\pi$
radians, hence:

\begin{displaymath}
({1 \over R}-R) \, \tilde \phi_{{\rm A,ion}}^{{\rm low}} \approx ({1 \over
  R}-R) \, \tilde \phi_{{\rm B,ion}}^{{\rm low}} 
\end{displaymath}

and {\it c)}, that the instrumental phase errors, due to independent
frequency standards and electronic equipment at each antenna, are
common for the observations of the target and calibrator sources,
then:

\begin{equation}
\phi_{{\rm A,inst}}^{{\rm high}} - R \,.\, \tilde \phi_{{\rm A,inst}}^{{\rm low}} \approx 
\phi_{{\rm B,inst}}^{{\rm high}} - R \,.\, \tilde \phi_{{\rm B,inst}}^{{\rm low}}  .
\eqnum{6}
\end{equation}

This is described in more detail in Section \ref{sec:error}.2 \\

These conditions are far less restrictive than the ones that apply for
conventional PR techniques, and can be easily met using
source-switching cycles up to $\sim \,$several minutes, and switching
angles up to $\sim \,$several degrees for SFPR observations at high
frequencies (further details are to be found in Section \ref{sec:guide}).

Making use of these approximations,  the 
{\sc SFPR}-calibrated visibility phases of the target dataset,
$\phi^{{\rm SFPR}}_{{\rm A}}$, are:

\eqnum{7}
\begin{eqnarray}
  \phi^{{\rm SFPR}}_{{\rm A}} 
  = \phi_{{\rm A,str}}^{{\rm high}} + 2 \pi \vec{D}_{\lambda^{{\rm high}}} \, . \, (\vec{\theta}_{{\rm A}}-
  \vec{\theta}_{{\rm B}}) + \Delta_{{\rm i,T^\nu_{\rm swt}}}  + \Delta_{{\rm i,T_{\rm swt}}}  ,
\end{eqnarray} 

\noindent
where $\Delta_{{\rm i,T_{\rm swt}}}$ stands for the interpolation errors arising
from using a source switching cycle ${\rm T}_{{\rm swt}}$.  The {\sc
  sfpr}-calibrated phases are free of tropospheric, ionospheric
and instrumental corruption, while keeping the chromatic astrometry
signature.  The terms $2\pi \,
\vec{D}_{\lambda^{{\rm high}}} \,.\, \vec{\theta}_{{\rm A}}$ and $2\pi \,
\vec{D}_{\lambda^{{\rm high}}}\,.\, \vec{\theta}_{{\rm B}}$ modulate the phase residuals
for each baseline with $\sim 24$ hours period sinusoid whose amplitude
depends on the magnitude of the ``core-shifts'' in ${\rm A}$ and ${\rm B}$,
respectively.  It is interesting to note that the ``core-shift''
functional dependence in Eq. 7 is identical to that for the pair
angular separation in conventional PR, although
the latter can not be applied at high
frequencies.  The propagation of interpolation errors into the SFPR
analysis is addressed in the error analysis in Section \ref{sec:error}.

\subsection{\it Outcomes of SFPR}\label{sec:m_astrometry}

\noindent
This section is concerned with the outcomes of the SFPR technique:
high sensitivity source maps and precise astrometric measurements at the
highest frequencies.  The {\sc sfpr}-calibrated target dataset (see
Eq. 7) is Fourier inverted and deconvolved to yield a synthesis image
of the target source at the target frequency ($\nu^{{\rm high}}$). We
call this the {\sc sfpr}-map.  The effective coherence time of the
{\sc sfpr}-calibrated dataset is increased as a result of the
tropospheric calibration derived with fast-frequency switching
observations, and further improved after the ionospheric and
instrumental calibration using the observations of a calibrator
source.  The result is a lower detection threshold for the SFPR
observations at $\nu^{{\rm high}}$.  This technique is therefore suitable
for the detection of weak sources, even at the high frequencies beyond
the applicability of conventional phase referencing techniques as demonstrated in \citet{middelberg_05_fs}.

SFPR observations on their own enable {\it bona-fide} ``chromatic''
astrometry, that is, measurements of frequency-dependent source
position shifts.  Eq. 7 shows that the {\sc
  sfpr}-calibrated phases are sensitive to ``core-shifts'' in the
target and the calibrator sources, which can be measured directly in
the {\sc sfpr}-map.  Given the similarity between the functional
dependence of ``core-shifts'' and source pair separation in the calibrated
datasets using SFPR and PR techniques, respectively, it is easy to see
that these can be measured in the {\sc sfpr}-map in a
similar way as angular separations are measured in PR maps.  That is, the position offset of the target
source relative to the {\sc sfpr}-map center corresponds to a {\it
  bona-fide} high precision astrometric measurement of the relative
``core shifts'' in the calibrator and target sources between the two
observed frequencies.

The combination of SFPR observations, at $\nu^{{\rm low}}$ and $\nu^{{\rm high}}$,
and PR observations at $\nu^{{\rm low}}$,
enables relative astrometry (i.e. measurement of the target source
position with respect to an external reference frame) at $\nu^{{\rm high}}$.
Therefore it is possible to achieve relative astrometric measurements
at the highest frequencies, beyond the $\sim$43-GHz upper limit in
conventional PR techniques. The requirements for the switching
cycles and the calibrator source in SFPR observations are relatively
easy to fulfill, hence are compatible with a wide range of targets and
applications.  Performing multi-epoch observations will enable high
precision parallax and proper motion studies at the highest
frequencies.

\section{Error Analysis}\label{sec:error}

This section is concerned with the propagation of errors in the VLBI
observables listed in Eq. 1, and from the interpolation process, into
the astrometric estimates using SFPR calibration techniques.  To
simplify the presentation, the individual sources of errors are
grouped into subsections 3.1 and 3.2 according to their dispersion
properties, or equivalently, to their compensation within the {\it
  Step 1} or {\it Step 2} interpolation process in the SFPR
calibration strategy. The third subsection 3.3 is devoted to
the thermal noise error contribution.

The propagation of the spatial and temporal interpolation errors in
the data analysis using conventional PR techniques has
been studied thoroughly using analytical and simulation studies
\citep{shapiro_79,beasley_pr,fomalont_95,pradel_06,asaki_pr}.
Here, we present an analytical study of the errors in SFPR techniques
based on a modified version of the error analysis for conventional
PR techniques presented in Asaki \etal\ (2007) ({\it
  hereafter} A07).

\subsection{\it Non-dispersive Terms: Tropospheric and Geometric errors (Step 1)}

The dominant contribution to the VLBI tropospheric errors arise from
inadequately modeling the inhomogeneous and highly variable
distribution of water vapor content, the so-called {\it wet} part, in
the troposphere.  Typical values of the equivalent tropospheric zenith
excess path delay are $\sim$3--5\,cm, for each telescope, and have a
non-dispersive nature.  The VLBI tropospheric errors exhibit random or
semi-random temporal fluctuations, which limit the coherence times,
and quasi temporally-invariant spatial variations along different
sight line directions.  They are referred to as the {\it dynamic} and
{\it static} components of the troposphere, respectively.

To correct for the VLBI tropospheric errors, and in general for any
errors, conventional PR uses fast source-switching observations and
the interpolation between scans of a calibrator source to the target
source scans, which is along a different line of sight.  After
correcting for the interpolated residual values the remaining errors
in the target dataset are attenuated, by a factor which is inversely
proportional to the source pair angular separation, but can still be
significant. At frequencies above $\sim$10\,GHz, the dominant source
of errors in the astrometric estimates are the uncertainties in the
tropospheric delay model, even when using a nearby ($\sim 1^o$ away)
calibrator source as shown in the simulation work of \cite{pradel_06}
and A07.  Using simultaneous observations of the sources, i.e. with
VERA \citep{vera}, results in an improved compensation of the errors
arising from the {\it dynamic} component of the troposphere, but those
from the {\it static} component still remain, and pose a limit in the
achievable astrometric precision, irrespective of the observing
frequency.  In order to achieve the very highest astrometric precision
additional independent measurements to reduce the {\it static} tropospheric
errors are required \citep{reid_04, honma_08}.
 
Instead, the SFPR technique {\it Step 1} corrects the tropospheric
errors in the target dataset using fast frequency-switching
observations of the same source 
plus interpolation between 
scans at the reference frequency to calibrate the 
target frequency scans, along the same line of sight,
after scaling by the frequency ratio.

Having observations along the same line-of-sight offers considerable
advantages for the precise calibration of the {\it static}
tropospheric errors, as shown below.  Additionally, in the ideal case
of simultaneous dual frequency observations the need for interpolation between
consecutive scans at the reference frequency is eliminated and an
exact calibration of the {\it dynamic} component of the troposphere can be
achieved as well.

Taking into account the similarity between the two calibration
techniques, we propose to use a modified version of the formulae in
A07 for conventional PR, to characterize the propagation of
non-dispersive errors in SFPR.  These modifications of the formulae
consist of {\it a)} replacing the source switching cycle in PR by the
frequency switching cycle in SFPR, {\it b)} replacing the source pair
angular separation in PR, by the ``core-shift'' values between the two
frequencies in SFPR, {\it c)} adding a multiplicative factor equal to
the SFPR frequency ratio {\it R} to give the phase errors at the
target frequency ($\nu^{{\rm high}}$), whilst using the reference frequency
($\nu^{{\rm low}}$) in the formula, and {\it d)} adding a multiplicative
factor of $\sqrt{2}$ to adapt for ground-ground rather than
space-ground baselines, as used in A07.

By doing this, the estimated SFPR residual phase error at the target
frequency ($\nu^{{\rm high}}$)
due to the the {\it   dynamic} component of the troposphere, $\sigma \phi^{{\rm high}}_{{\rm dtrp}}$, for
dual frequency observations of source ${\rm A}$, with a given baseline,
becomes:

\eqnum{8}
\begin{eqnarray}
\sigma \phi^{{\rm high}}_{{\rm dtrp}} [{\rm deg}] \approx \sqrt{2} 
R \, 27 \, C_w \, \left({\nu^{{\rm low}}[{\rm GHz}] \over 43 {\rm GHz}}\right)
\left({\sec{Z_g} \over \sec{45^o}}\right)^{1/2} 
\times  \left({\rm {T^{\nu}_{swt} [s]} \over {\rm 60s}} + 
0.16 \left({\sec{Z_g} \over \sec{45^o}}\right) \left({\theta_A [^o] \over 2^o}\right)\right)^{5/6}
\nonumber \\
\approx 
R \, 38 \, C_w \, \left( {{\rm \nu^{low}}[{\rm GHz}] \over 43 {\rm GHz}} \right)
\left( {\sec{Z_g} \over \sec{45^o}}
\right)^{1/2} 
\times  \left( {{\rm T^{\nu}_{swt}} [{\rm s}] \over {\rm 60s}} \right)^{5/6}, 
\end{eqnarray}

where $\nu^{{\rm low}}$ is the reference frequency, ${\rm T^{\nu}_{swt}}$ is the
frequency switching cycle, $C_w$ is a modified coefficient of the
troposphere spatial structure function to characterize the weather
conditions (with values 1, 2, and 4 for good, typical and poor
tropospheric conditions, respectively), and $Z_g$ is the telescope
zenith angle. $\theta_A$ stands for the magnitude of the
``core-shift'' between $\nu^{{\rm low}}$ and $\nu^{{\rm high}}$ which
for AGNs for example, has
typical values of a few hundreds of $\mu$as at centimeter wavelengths, and 
is expected to be reduced at higher frequencies. 
The random phase noise introduced by the short-term tropospheric fluctuations 
is expected to affect the quality of the map by producing blurred images, 
but not shifts of the peak of brightness that lead to astrometric errors. 
Eq. 8 shows that such turbulence errors 
can be attenuated using fast frequency switching cycles. Ultimately, note
that these errors become zero in the case of simultaneous
(${\rm T^{\nu}_{swt}}$=0) dual frequency observations. 
Table \ref{tab:one} lists the estimated $\sigma \phi^{{\rm high}}_{{\rm dtrp}}$
residual phase values using both fast frequency switching and
simultaneous dual frequency observations for two pairs of frequencies,
namely 43/86 GHz and 43/129 GHz, with typical observing parameters for
SFPR; for comparison, the estimated values
for conventional PR at 43 GHz are also listed.\\
 
Similarly, the SFPR residual errors at $\nu^{{\rm high}}$ due to the 
{\it static} component of the troposphere, $\sigma 
\phi^{{\rm high}}_{{\rm strp}}$, are expressed as: 

\eqnum{9} 
\begin{eqnarray}
\sigma \phi^{{\rm high}}_{{\rm strp}} [{\rm deg}] \approx \sqrt{2}\, R \, 76 \left({{\rm \nu^{low}} [{\rm GHz}] \over 43
 {\rm GHz}}\right) \left({\Delta l_z [{\rm cm}] \over 3 {\rm cm}}\right) 
\left({\theta_A [^0] \over
  2^o}\right) \left({\cos{Z_g} \over \cos{45^o}}\right)^{-1} \left({\tan{Z_g} \over \tan{45^o}}\right) 
\approx 0 , 
\end{eqnarray}

where $\Delta l_z$ is the uncertainty in the tropospheric zenith
excess path length (typical values, $\sim$\,3--5 cm). 
All the systematic (or {\it static}) errors are strongly attenuated by
the small magnitude of the ``core-shift'' ($\theta_{\rm A}$) parameter.  For comparison,
the attenuation factor resulting from a typical AGN core-shift value in SFPR
observations is more than $10^7$ times greater than that in PR
observations of a pair of sources with $\sim 1^o$ separation.
These systematic errors
set a barrier in the astrometric precision achievable with
conventional PR techniques, as mentioned above.  Instead, they are
completely suppressed in SFPR techniques due to the same
line-of-sight dual frequency observations (except for the effectively
zero angular separation due to the ``core-shift''). 
Table \ref{tab:one} lists the estimated $\sigma \phi^{{\rm high}}_{{\rm strp}}$
residual phase values for typical observing parameters with SFPR
observations; for comparison, the estimated values using conventional
PR at 43 GHz are included.

For the same reason the errors arising from any inadequacies in 
the ``a priori'' geometric delay model, comprising errors in the
source position ($\sigma \phi_{{\rm \Delta s}}$) and antennae
coordinates ($\sigma \phi_{{\rm bl}}$), among others, are
also readily compensated in {\it Step 1}.

\begin{displaymath}
\sigma \phi^{{\rm high}}_{{\rm \Delta s}} [{\rm deg}] \approx R \, 16
\left({\nu^{low} [{\rm GHz}] \over 43 {\rm GHz}}\right)
\left({\rm B [km] \over 6000 km}\right) \left({\Delta s^c [{\rm mas}] \over 0.3 {\rm mas}}\right)
\times \left({\theta_{\rm A} [{\rm deg}] \over 2^o}\right)
\approx 0  ,\\
\end{displaymath}

\begin{displaymath}
\sigma \phi^{{\rm high}}_{{\rm bl}} [{\rm deg}] \approx R \, 18
\left({\nu^{{\rm low}} [{\rm GHz}] \over 43 {\rm GHz}}\right)
\left({\rm \Delta P [cm] \over 1 cm}\right)
\times \left({\rm \theta_A [deg] \over 2^o}\right)
\approx 0  , \\
\end{displaymath}

where ${\Delta s^c}$ is the source position error, ${\rm B}$ is the
projected baseline length, and $\Delta {\rm P}$ represents the combined
contribution from the Earth Orientation Parameters and both the
antenna coordinate errors. Table \ref{tab:one} lists the estimated
values for $\sigma \phi^{{\rm high}}_{{\rm geo}}$ which is comprised of the terms
above.

Summarizing, the propagation of systematic tropospheric errors 
({\it static} component) into the estimates using
SFPR calibration techniques is negligible because of the attenuation
from the near-identical lines of sight for the two observing
frequencies, and the same applies to geometric errors.
Furthermore the short-term tropospheric errors ({\it dynamic}
component) will effectively cancel with simultaneous
dual frequency observations, since ${\rm T^{\nu}_{swt}=0}$.  Therefore, the
capability for simultaneous multiple frequency-band observations, with
arrays like KVN and telescopes like Yebes and Haystack, gives an exact
adaptive tropospheric correction using SFPR techniques.
The advantage of this being an extended coherence time (at the target
frequency), so a greater number of weaker sources will be detectable
by means of longer integration times, as well as the improved quality
of the {\sc sfpr}-map and more accurate astrometric measurements.

\subsection{\it Dispersive Terms: Ionospheric and Instrumental errors 
 (Step 2) }

The inaccuracies in the ionospheric delay model introduce a dispersive
phase error which is inversely proportional to the observing
frequency.  The effect of the ionospheric errors is conveniently
described as a compound of {\it static} and {\it dynamic} components,
similar to the description for the troposphere.  The latter introduces
temporal phase fluctuations caused by the irregularities in the plasma
density in the ionosphere; the former introduces quasi temporally
invariant spatial variations arising from uncertainties in the
vertical total electron content (TEC) and the geometry of the
atmosphere.

Although the ionospheric effects at the high frequencies of interest
for this paper are expected to
be weak, and are certainly much smaller than the tropospheric
errors, they still need to be accounted for. Attempts to achieve
astrometry using only fast-frequency switching strategies 
failed because those were not corrected for \citep{middelberg_05_fs}.

Calibration {\it Step 2} of the SFPR technique corrects the
tropospheric-free target dataset for ionospheric errors by using
interleaved observations of an external calibrator source, and
interpolation to the target source scans, following a similar strategy
as in conventional PR.  Despite its similarity to PR, the constraints
on the source switching cycle and the angular separation between the
sources are much less strict in SFPR due to the preceding tropospheric
calibration ({\it Step 1}).  It should be noted that a side effect of
the tropospheric calibration is that the effective residual
ionospheric errors at the target frequency are those at the reference
frequency, $\nu^{{\rm low}}$, multiplied by a factor $R-1/R$ ({see Eq. 3});
hence the ionospheric time and space coherence (i.e. the ionospheric
isoplanatic patch size) are effectively reduced by a similar
factor. However these are still much less significant than the
dominant tropospheric errors in conventional PR.

Taking this into account, we propose a modified version of the A07
formula that describes the interpolated SFPR residual phase errors at the
target frequency ($\nu^{{\rm high}}$) introduced by
the {\it dynamic} ionosphere, $\sigma \phi^{\nu^{{\rm high}}}_{{\rm dion}}$:

\eqnum{10} 
\begin{eqnarray}
\sigma \phi^{{\rm high}}_{{\rm dion}} [{\rm deg}] \approx \sqrt{2} \left(R-1/R\right) \, 0.46 \, 
\left({\sec{Z_i} \over \sec{43^o}}\right)^{1/2}
\left({\nu^{{\rm low}} [{\rm GHz}] \over 43 {\rm GHz}}\right)^{-1} \nonumber \\
\times \left[0.21 \left({\rm T_{swt} [s] \over  60 s}\right) +
\left({\sec{Z_i} \over \sec{43^o}}\right) \left({\Delta \theta [^o] \over 2^o}\right)\right]^{5/6}  , 
\end{eqnarray}

where ${\rm T_{swt}}$ is the source-switching cycle between the target and
calibrator sources, $\Delta \theta$ is their angular separation
(i.e. switching angle), and $Z_i$ is the zenith angle measured at an
altitude of $\sim 300\,$ km (bottom of the ionospheric F-region) which
corresponds to the height of the phase screen model of the ionospheric
propagation effects.  Table \ref{tab:one} lists the estimated $\sigma
\phi^{{\rm high}}_{{\rm dion}}$ residual phase values, for observations at
43/86GHz and at 43/129 GHz, using two source pair angular separations,
equal to 2 and 10 degrees, and a source-switching cycle of 5 minutes.
Based on these estimates, the effect of the residual ionospheric
random phase noise (introduced by the short-term ionospheric
fluctuations) on the quality of the {\sc sfpr-}ed map is negligible at
the high frequencies of interest for this paper.

Similarly, the interpolated SFPR residual phase errors arising from the 
{\it static} component of the ionosphere can be expressed as:

\eqnum{11}
\begin{eqnarray}
\sigma\phi^{{\rm high}}_{{\rm sion}} [{\rm deg}] \approx \sqrt{2} \left(R-1/R\right)\, 2.7 \, 
\left({\nu^{{\rm low}} [{\rm GHz}] \over 43 {\rm GHz}}\right)^{-1} 
\left({\Delta I_V [{\rm TECU}] \over 6 {\rm TECU}}\right) 
\left({\Delta \theta [{\rm deg}] \over 2^o}\right) \nonumber \\
\times \left({\cos{Z_F} \over \cos{41^o}}\right)^{-1} \left({\tan{Z_F} \over \tan{41^o}}\right) , 
\end{eqnarray}

where $\Delta I_{\nu}$ is the systematic error in the estimate of the
vertical TEC, TECU is the TEC unit (10$^{16}$
electrons\,m$^{-2}$), and $Z_F$ is the zenith angle measured at the
altitude of the electron density peak (typically, 450 km).

The propagation of long-term systematic ionospheric errors is
attenuated by a factor proportional to the source pair angular
separation, expressed in radians, and its value can be large for wide
switching angles.

Such long-term phase errors can distort the image and affect the
astrometric measurements in the {\sc sfpr-}ed map. In SFPR
observations, which provide a quasi perfect compensation of systematic
tropospheric errors, the ionospheric errors can become the dominant
source of astrometric errors.

Table \ref{tab:one} lists the estimated $\sigma \phi^{{\rm high}}_{{\rm sion}}$ 
residual phase values, for observations at 43/86GHz and at 43/129 GHz, 
using  source pair angular separations of 2 and 10 degrees, 
and a source-switching cycle of 5 minutes.
Based on these estimates, at the high frequency regime of interest in this
paper, source-switching cycles of several minutes, and switching
angles up to 10 degrees, or even larger, are acceptable for SFPR observations.

The residual instrumental phase errors in Eq. 2 are also compensated
in the {\it Step 2} calibration, along with the ionospheric errors.  These
contributions arise from the excess
path introduced by the electronics and independent frequency standards
and elevation-dependent structure deformations at each antenna.
Albeit difficult to estimate, such contributions are expected to be
slowly varying with time, as a result of changes of ambient conditions
and with observing elevation angles.  
Therefore, a slow source switching
cycle of several minutes, as required for the compensation of
ionospheric errors, is suitable to compensate for the instrumental
errors as well
(i.e. $\phi_{{\rm A,inst}}^{{\rm high}} \approx \phi_{{\rm B,inst}}^{{\rm high}} \,\,;\,\,
\tilde \phi_{{\rm A,inst}}^{{\rm low}} \approx \tilde \phi_{{\rm B,inst}}^{{\rm low}}$), and
meet the working condition for {\it Step 2} expressed in Eq. 6. 
Hence, the instrumental errors in SFPR $\sigma \phi_{{\rm inst}}$ are negligible:

\begin{displaymath}
\sigma \phi_{{\rm inst}} = \left(\phi_{{\rm A,inst}}^{{\rm high}} - R \,.\, \tilde \phi_{{\rm A,inst}}^{{\rm low}}\right) -
\left(\phi_{{\rm B,inst}}^{{\rm high}} -  R \,.\, \tilde \phi_{{\rm B,inst}}^{{\rm low}}\right) \approx 0 .
\end{displaymath}

\subsection{\it Thermal Noise}

The {thermal noise} in the VLBI observables sets the final
insurmountable barrier to the map quality and the astrometric
precision, in the absence of other sources of errors.

The general expression for the error in the phase measurements 
with a pair of antennas {\it i} and {\it j} due to thermal noise is given by:

\begin{displaymath}
 \sigma \phi_{{\rm thermal}} [{\rm rad}] = \Delta S_{i,j}/S = {1 \over \eta_s}
\sqrt {{\rm SEFD_i [Jy]} \times
  {\rm SEFD_j [Jy]} \over 2 \Delta \nu {\rm [Hz] T [sec]}} 
{1 \over S [{\rm Jy}]}  , 
\end{displaymath}

from \cite{walker_95}, where $\Delta S_{i,j}$ is the detection
threshold for the baseline and $S$ is the source flux. 
${\rm SEFD= T_{sys} / G}$ (with ${\rm T_{sys}}$ the
system temperature, and ${\rm G}$ the antenna gain) is the antenna's system
equivalent flux density, a parameter that measures the overall
performance of each antenna, $\eta_s$ is the interferometer system
efficiency which accounts for digital losses, ${\rm T}$ is the integration
time, and $\Delta \nu$ is the bandwidth collected at each antenna, 

An equivalent expression for two-bit data sampling, as given by A07, is: 

\eqnum{12}
\begin{equation}
\sigma \phi_{{\rm thermal}} [{\rm deg}] = 1.6 \times 10^{-5} \, \left({\Delta \nu
[{\rm MHz}] \over 256 {\rm MHz}}\right)^{-1/2} \, \left({\rm T [s] \over 10 s}\right)^{-1/2} \,
\left({ \overline{{\rm SEFD}} [{\rm Jy}] \over S [{\rm Jy}]}\right) ,
\end{equation}

where $\overline{{\rm SEFD}}$ stands for the geometric mean of the ${\rm SEFD}$
values for the two antennas.

An expression for the thermal noise phase error resulting from the
{\it Step 1} calibration in SFPR techniques can be
written as:

$\sigma \phi^{{\rm \nu^{low}}, \nu^{{\rm high}}}_{{\rm A,thermal}} = 
\sqrt {[\sigma \phi^{{\rm \nu^{high}}}_{{\rm A,thermal}}]^2 + {R^2 \over 2}  
[\sigma \phi^{{\rm \nu^{low}}}_{{\rm A,thermal}}]^2}$ ,

where $\sigma \phi^{{\rm \nu^{high}}}_{{\rm A,thermal}}$ and $\sigma
\phi^{{\rm \nu^{low}}}_{{\rm A,thermal}}$ stand for the thermal noise error in the
phase measurements at the target ($\nu^{{\rm high}}$) and reference
($\nu^{low}$) observing frequencies, respectively, of source ${\rm A}$,
as given by Eq. 12. The factor $\sqrt{{R^2 \over 2}}$, where
$R=\nu^{{\rm high}} / \nu^{{\rm low}}$, comes from the scaling and interpolation
operations between consecutive reference frequency scans to calibrate
the interleaving scan at the target frequency, on a common source.
An identical expression applies for the observations of source ${\rm B}$, 
$\sigma \phi^{\nu^{{\rm low}}, \nu^{{\rm high}}}_{{\rm B,thermal}}$. 

Similarly, an expression for the thermal noise phase error 
resulting from the {\it Step 2} calibration can be written as:

$\sigma \phi^{\nu^{{\rm high}}}_{{\rm A, B,thermal}} = 
\sqrt {[\sigma \phi^{\nu^{{\rm high}}}_{{\rm A, thermal}}]^2 + {1 \over 2}  
[\sigma \phi^{\nu^{{\rm high}}}_{{\rm B,thermal}}]^2}$ , 

where $\sigma \phi^{{\rm \nu^{high}}}_{{\rm A, thermal}}$ and $\sigma
\phi^{\nu^{{\rm high}}}_{{\rm B, thermal}}$ are given by Eq. 12, and the factor
${1 \over \sqrt{2}}$ results from the interpolation between
consecutive calibrator source (${\rm B}$) scans to calibrate the target
source (${\rm A}$) scans in-between.  Note that the integration time in {\it
  Step 2} calibration is extended beyond the nominal coherence time at
$\nu^{{\rm high}}$ as a result of the preceding tropospheric calibration.

Finally, the thermal phase noise contribution using SFPR techniques,
for a baseline, can be expressed as the root sum square of the 
contributions above, that is: 

$\sigma \phi^{{\rm high}}_{{\rm SFPR,thermal}} = 
\sqrt { (\sigma \phi^{\nu^{{\rm low}}, \nu^{{\rm high}}}_{{\rm A,thermal}})^2 + 
(\sigma \phi^{\nu^{{\rm low}}, \nu^{{\rm high}}}_{{\rm B,thermal}})^2 + 
(\sigma \phi^{\nu{{\rm high}}}_{{\rm A,B,thermal}})^2}$ .

Taking typical SEFD parameter for the VLBA antennas at 43 and 86 GHz,
and source fluxes equal to 0.1\, Jy with a bandwidth of 256 MHz and
integration time of 10\,seconds, the value of $\sigma
\phi_{{\rm SFPR,thermal}}$ is $\sim \, 1.5^o$ at 86 GHz.  In general, this
contribution is insignificant compared to other sources of errors in
the astrometric analysis.  Nevertheless using SFPR techniques, with
simultaneous dual-frequency observations, one would approach the thermal
limits at sub-millimeter wavelengths.

\section {Guidelines for Scheduling SFPR Observations}\label{sec:guide}

This section provides practical guidelines for scheduling SFPR
observations from a perspective of minimizing the analysis errors.

\subsection{\it Integer frequency ratio and magnitude of R} 

In general, it is strongly recommended that the two frequencies
involved in the SFPR technique have an integer ratio in order to avoid
phase-ambiguity related problems in the analysis.  Such
phase-ambiguity issues arise from the inherent unknown number of
$2\pi$ cycles in the measured phase values, as shown in Eq. 2.  The
scaling by the frequency ratio $R$ involved in the {\it Step 1}
calibration of SFPR will continue to keep this unknown term as a whole
number of cycles, and transparent to the analysis, as long as the
scaling factor is an integer number.  Note that it is sufficient that
the recorded bandwidths cover the frequencies with an integer ratio. 
Preliminary results from our simulations suggest that 
extrapolation to a frequency value outside the covered band might be 
feasible in certain cases, nevertheless further investigation is
required and results will be reported somewhere else.
Non-integer frequency ratios
will, in general, introduce phase offsets and jumps which would have
to be addressed separately. The exception to that being the rare case
when the number of phase turns is the same for both sources, as for
example in observations of the extremely nearby quasar pair 1038+528 A
and B, 33$^{\prime \prime}$ apart as demonstrated in
\citet{rioja_05_cs}.

The value of $R$, besides being integer, has an impact on the
ionospheric phase compensation, as shown in Eq. 3.  For a given target
frequency, the magnitude of the residual ionospheric phase errors will
increase with larger values of $R$, or equivalently lower reference
frequency. The reason for this is twofold, first, due to the increased
ionospheric effects at lower reference frequencies, and second, due to
the multiplicative factor ($R-1/R$). Instead, the magnitude of the
non-dispersive errors is independent of the value of
$R$.  As for the thermal noise errors the combined
effect of $R$, the receiver noise (SEFD) and the source fluxes must
be considered on a case-to-case basis, but are generally much smaller
than the ionospheric effect.

Therefore, the advantages of using as the reference frequency a lower
frequency, such as better receiver performance, higher intrinsic
source flux and longer coherence times, should be balanced against the
issues raised by the propagation of increased ionospheric effects at
lower frequencies, further amplified by the $R-1/R$ factor in the SFPR
analysis.  As a compromise, in absence of other reasons to choose the
observing frequencies, we generally recommend the use of a reference
frequency equal to 22 GHz or above (whilst avoiding 22.35\,GHz where
water absorption is problematic), where ionospheric effects start to
be weak and the receiver performance is good. 

\subsection{\it Switching times and Switching angle}

The dual frequency observations are the basis of the ``{\it Step 1}''
calibration in the SFPR technique. They enable an
adaptive phase-based calibration of the tropospheric fluctuations 
at the target frequency using fast frequency switching  
observations of a common source. 
The frequency switching cycle has to be fast enough to fully sample
the tropospheric fluctuations and ensure unambiguous phase connection
between consecutive scans at the reference frequency.  Without a
reliable phase connection it is not possible to proceed any further in
the calibration therefore this is the most critical consideration in
the scheduling.

At frequencies $\ge \,10$-GHz the temporal variations in the residual
phases are dominated by the random fluctuations that arise in the
dynamic component of the troposphere.  Therefore, the dynamic
troposphere is the determining factor on setting the
frequency-switching cycles at the frequencies of interest for
SFPR.  As a rule of thumb the same guidelines for telescope switching
cycles in conventional PR at the reference frequency $\nu^{{\rm low}}$
(see \cite{beasley_pr,memo_20}), including the weather dependence,
apply for the frequency-switching cycles with SFPR. 
Those range between one and several minutes for observations at
22 GHz under typical and good weather conditions; for 43 GHz, 
between several tenths of a minute to a few  minutes.
Also, the switching cycle has to allow for direct detections of both sources at
the reference frequency, as explained below in subsection 4.3.

Having a fast frequency switching cycle reduces the magnitude of
(random) tropospheric errors, as shown in Eq. 8, which result in a
blurring effect.  Ultimately, by using simultaneous observations at
multiple frequencies, as the KVN will do, this error term will vanish
completely.  Other advantages are that since a tropospheric
interpolation would not be required phase connection is not an issue,
and the increased on-source time means that weaker sources would be
suitable for SFPR observations.

The source switching strategy is the basis of the {\it Step 2}
calibration in SFPR techniques to compensate for remaining ionospheric
and instrumental effects, after the tropospheric effects have been
removed.  The ionospheric effects are weak and the isoplanatic patch
size is large at high frequencies, therefore source switching cycles
of several minutes and large angular separations are acceptable.
However, the ionospheric isoplanatic patch size is difficult to
characterize exactly, as it is very variable in time and in Earth
location. Nevertheless if the ionosphere is smooth, the calibrator
source should not need to be close, certainly not as close as for
PR. Extrapolating from PR astrometric errors, a $10^o$ or even $20^o$
ionospheric patch size is possible.  Therefore, unlike as in PR, the
conditions for finding a suitable SFPR calibrator source are much less
restrictive.

Since the calibrator scans reduce the observing time on 
the target source, we advise scheduling the observations to achieve a
reasonable (i.e. to the level of other error contributions)
ionospheric calibration, following Eqs. 10 and 11, with the minimum
calibration overhead.

\subsection{\it Target/Calibrator Source fluxes} 

A necessary condition for successful application of SFPR techniques is
that the calibrator and target sources must be directly detected at
the reference frequency scans (i.e. within the coherence time imposed
by the tropospheric fluctuations).  Typical minimum source flux values
are the same as for the calibrator source in conventional PR
observations at the reference frequency $\nu^{{\rm low}}$.  On the other
hand, direct detections at the target frequency $\nu^{{\rm high}}$ are not a
requirement, since the tropospheric calibration using dual-frequency
observations results in phase stabilization at the target frequency,
for both sources.  This enables increased sensitivity through use
of longer integration times (i.e. up to several minutes) in the
self-calibration of the troposphere-free calibrator dataset.  Finally,
the {\sc sfpr}-calibrated target dataset is expected to reach an
extended coherence time up to $\sim\,$hours, as in conventional PR.

The effect of the source fluxes into the final errors 
is described by the theoretical thermal noise limit (as shown in Eq. 12). 
Along with the source flux, other parameters that define this limit 
are the receiver noise (SEFD) and the frequency ratio $R$.
For example, to compare the thermal noise errors in SFPR observations
at 86 GHz either using 22 or 43 GHz as the reference frequency, one
should take into account the effect of halving $R$ compared to that of
the increased SEFD at 43 GHz.  For the current VLBA parameters one
needs to balance the 40\% reduction from the frequency-ratio
contribution against the approximately three-fold higher SEFD, for
43-GHz compared to 22-GHz.  Additionally one needs to account for the
fact that the sources tend to be weaker at higher frequencies.

\section{Observational demonstration}\label{sec:obs}

\subsection{\it Observations and Data Analysis}\label{sec:obs_ana}

On February 18, 2007, we carried out SFPR observations at 43 and 86 GHz,
using the eight antennas of the NRAO Very Long Baseline Array (VLBA)
which are equipped for the highest frequencies, for a total of 7 hours.
Based on the encouraging results from our error analysis, we selected 
as targets two pairs of well known and bright AGNs, with very
different angular separations: a close pair, 1308+326 \& 1308+328,
$14^\prime$ apart, and a widely separated pair, 3C273 \& 3C274 (M87),
$10^o$ apart. These source configurations provide two extreme cases of
application of SFPR techniques, and allow us
to investigate the effect of the source pair angular separation in our method.
Table \ref{tab:two} lists the theoretical error contributions estimated 
for these observations.

All sources were observed in the same run ({\em Exp. Code: BD119}),
alternating $\sim \,$1.5-hour long blocks on each of the source pairs
to improve the coverage of the {\it(u,v)}-plane.  For each source
pair, the observations consisted of alternating pointings
between the two sources (with a switching cycle of $\sim \,$5 minutes), 
and a rapid frequency-switching between 43 and 86 GHz (with a
switching cycle of $\sim \,$60 seconds). Hence, for a given source, 
the scan durations at each frequency were $\sim\,$20 seconds long, as the
time to change receivers at the VLBA is $\sim \,$10 seconds.
The observing switching cycles were chosen to match the
temporal scale of the ionospheric and tropospheric
fluctuations estimated for ``good'' weather conditions, respectively,
as requested for this experiment.
This results in a total on-source time equal to $\sim$30 minutes, for
each source at each frequency, after subtracting the receiver and
telescope slew times, and time devoted to pointing and other basic
calibration.

Additionally, a series of 1-hour long test observations ({\em
  Exp. Code: BD123}) were run on different days in February and March,
2007, using a similar observing schedule but with no weather
restrictions to test the robustness of the method weather-wise.  In
this case only the close pair of sources were observed because of time
constraints.  In all cases, each antenna recorded eight 16-MHz IF
channels, using 2-bit Nyquist sampling, which resulted in a data rate
of 512 Mbps.  

We used the NRAO AIPS package for the data reduction. We followed
standard VLBI calibration procedures for correcting for updated Earth
Orientation Parameters, GPS-measured TEC values, amplitude and
feed-rotation calibration (see \citet{memo_31} for details).
The astrometric analysis was carried out independently for each pair
of sources following a common two-step calibration strategy, as
explained in Section 2, with 1308+326 and 3C274 as target sources,
1308+328 and 3C273, as calibrator sources, and 86 and 43 GHz as the
target ({\it high}) and reference ({\it low}) frequencies,
respectively. First, we used the same-source observations at the two
frequencies to eliminate the dominant rapid tropospheric phase
fluctuations at 86 GHz, followed by a second calibration cycle to
correct for weaker, slower, remaining ionospheric and instrumental
contributions in the target data, using the observations of the
calibrator source.

Some details on the implementation of {\sc sfpr} using
AIPS are described here (see \cite{memo_31} for more information).
We used the AIPS task FRING to estimate the residual antenna-based
phases and phase derivatives (delay and rate) at 43 GHz, for each
source and for each scan of duration $\sim$20\,seconds, using a
point-source model.
The extended source structure contribution was removed using hybrid
maps as input models to CALIB, on the {\sc fring}-ed
data, for the 3C sources.
The SN-tables generated by the FRING and CALIB tasks provide the basis
for the subsequent calibration of the 86-GHz data sets, once the
estimated phase values have been scaled by 2 (i.e. the frequency
ratio).  The phase-scaling operation can be done with the SNCOR task,
using the function `XFER' (although it is meant for a different
purpose); for the delay and rate values, which are stored in
frequency-independent units of seconds and seconds/second
respectively, no changes are required.  Then, the entries in the
SN-tables were interpolated between consecutive 43-GHz scans, 60
seconds apart,
with the task CLCAL, and applied to calibrate the interleaving
observations of the same source at 86 GHz. The result of this
dual-frequency calibration are tropospheric-free (or {\sc
  fpt}-calibrated) datasets with increased coherence time at 86 GHz,
for both sources of each pair.  However the remaining errors
prevent the recovery of the position of the target source, using a
Fourier inversion, at this stage.

To recover the astrometry a further calibration iteration is required,
to disentangle the ``core-shift'' signature from the remaining
dispersive contributions in the tropospheric-free or {\sc
  fpt}-calibrated dataset
at 86-GHz, along with the source structure contribution.
To do this in AIPS, we re-FRING the {\sc fpt}-calibrated 1308+328
and 3C273 calibrator datasets at 86 GHz, and applied the interpolated adjusted
antenna phase, delay and rate solutions to the {\sc fpt}-calibrated
1308+326 and 3C274 target datasets at 86 GHz, respectively.
The resultant {\sc sfpr}-calibrated data-sets were Fourier inverted
and deconvolved with IMAGR, without further calibration, to produce the
{\sc sfpr}-maps for both target sources at 86 GHz.

\subsection{\it Observational Results}\label{sec:obs_res}

The application of SFPR techniques results in a {\sc sfpr}-ed map of
the target source, where the offset of the peak in the brightness
distribution with respect to the center of the map is astrometrically
significant.  As in conventional PR, the price to
pay for preserving the astrometric signature 
is that the quality of the reconstructed image is degraded as compared
to those in hybrid maps, due to residual calibration errors.  Nevertheless
these degraded images retain the astrometric information.

Table \ref{tab:two} lists the theoretically estimated {\it rms} phase error
contributions for our SFPR observations at 43/86 GHz of 2 pairs of
sources with very different angular separations, 14$^\prime$ and
$10^o$ respectively, using the VLBA. Note that the so-called {\it
  static} ionospheric errors are significantly higher for the pair
with larger separation, while the dominant {\it dynamic} tropospheric
errors, which depend on the frequency switching cycle and weather
conditions, are similar for both pairs.  All error values have been
calculated using the formulae from the error analysis (section 3) and
predict a successful outcome, albeit with reduced peak flux.
This is confirmed by the {\sc sfpr}-maps presented in this section,
which provide an empirical demonstration of the feasibility of SFPR
techniques for ``bona fide'' astrometry, even with very wide source
angular separations.

\begin{figure}
\includegraphics[width=\textwidth]{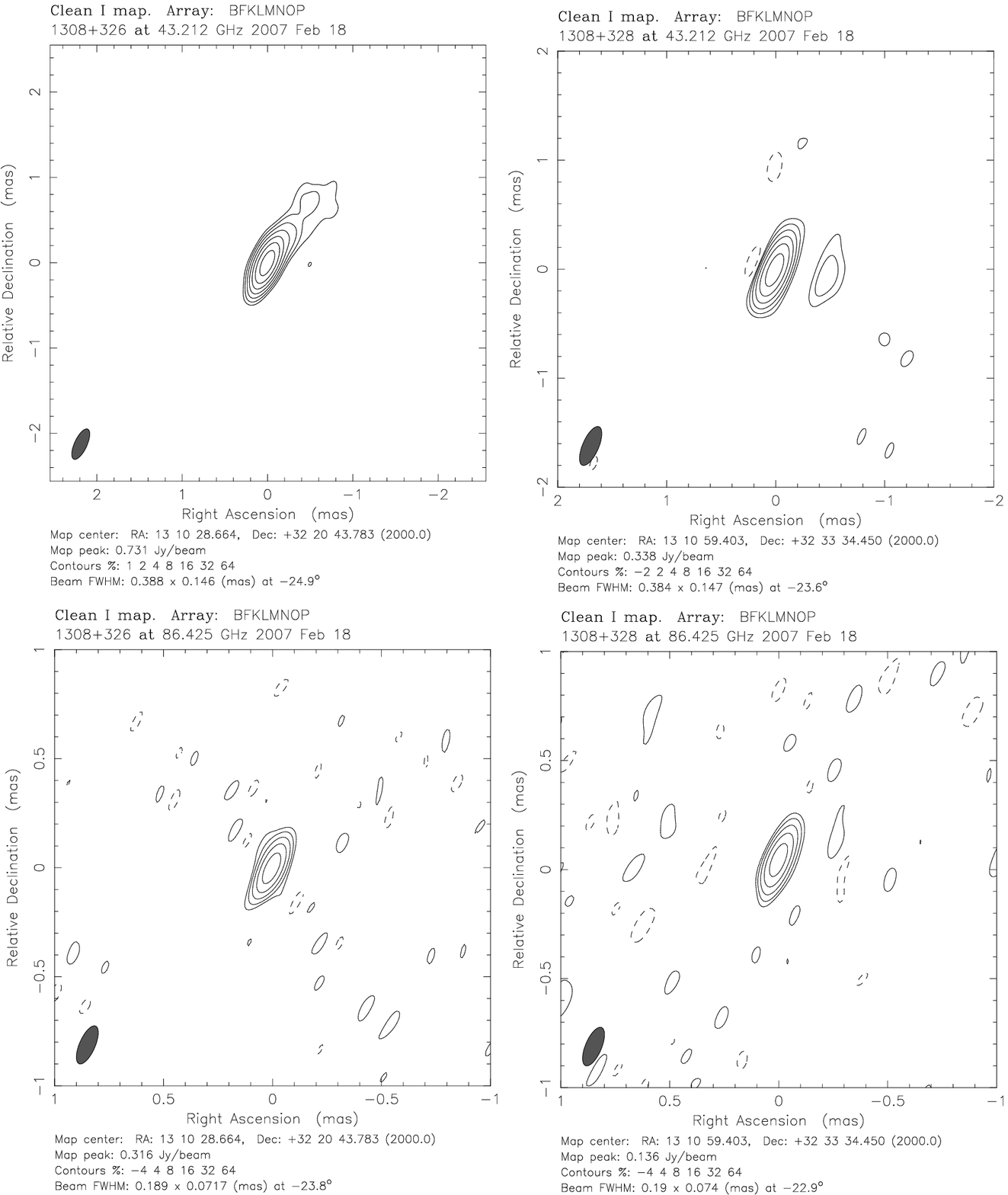}
\caption{Hybrid maps of 1308+326 (left) and 1308+328 (right) at 43
  (top) and 86 GHz (bottom). The data have been exported from AIPS and
  all antennas selfcalibrated and imaged with uniform weighting in difmap.}
\label{1308_hybrid}
\end{figure}

\begin{figure}
\includegraphics[width=\textwidth]{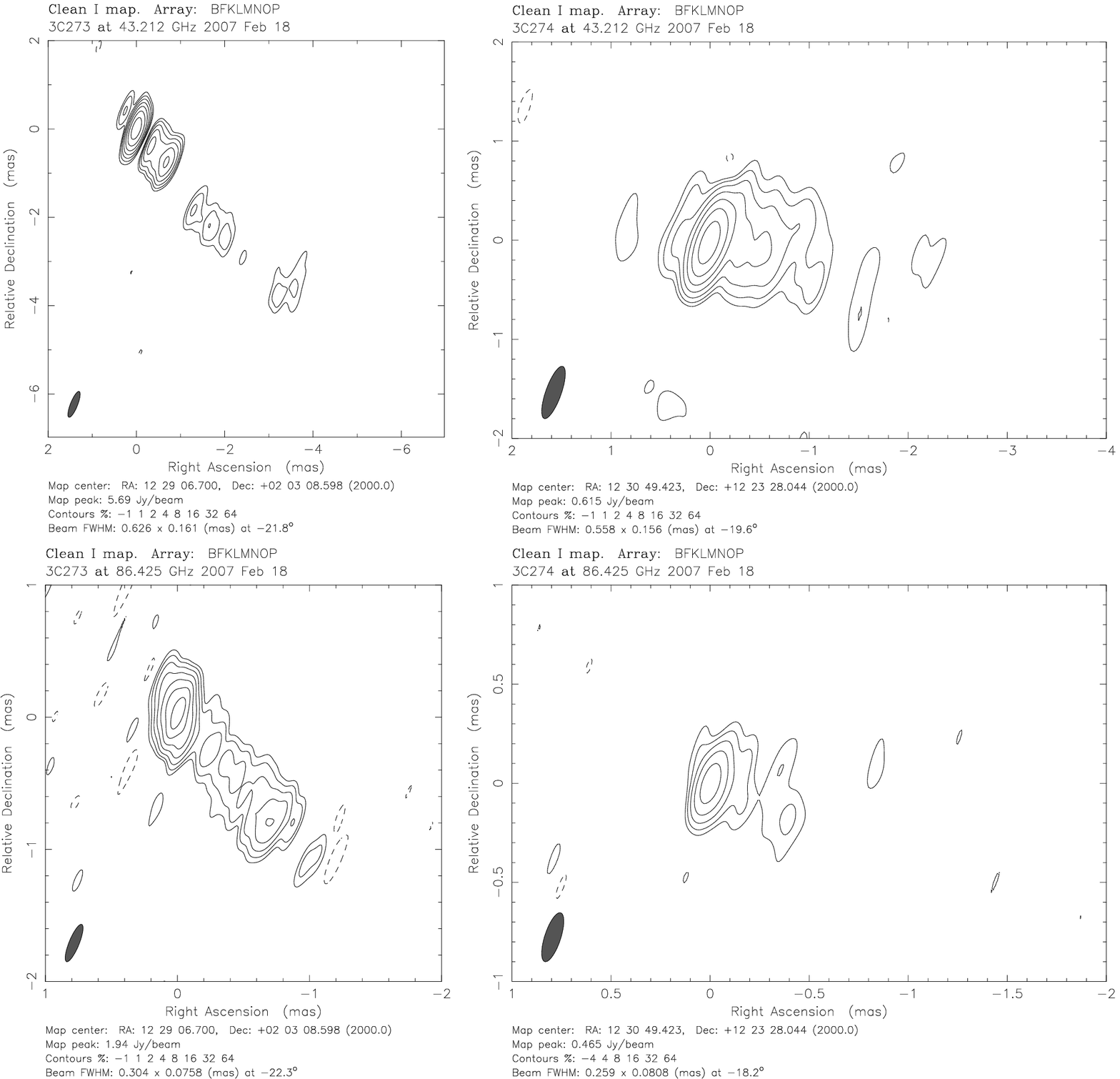}
\caption{Hybrid maps of 3C273  (left)  and 3C274  (right) at 43
  (top) and 86 GHz (bottom). The data have been exported from AIPS and
  all antennas selfcalibrated and imaged with uniform weighting in difmap. }
\label{3c_hybrid}
\end{figure}

Figures \ref{1308_hybrid} and \ref{3c_hybrid} show the hybrid maps for
all observed sources at 43 and 86 GHz. The peak flux from these maps
will be used for calculating the flux recovery quantity, or Strehl
ratio, which is defined as the ratio between the peak fluxes in the
{\sc sfpr}-maps and hybrid maps.  This relates to the variance of a
random Gaussian phase noise, $\sigma_{{\rm \phi}}$, as
$e^{-\sigma^2_{\phi}/2}$ \citep{thompson_isra_2}. Therefore it
provides an empirical estimate of the magnitude of the random errors
in the {\sc sfpr-}maps, and of the corresponding astrometric
\underline{precision}.  Nevertheless this approach is not sensitive to
systematic errors, which could potentially bias the astrometric
\underline{accuracy}, and further error analysis, as presented in
Section 3, is required to provide a reliable error estimate.

\begin{figure}
\includegraphics[angle=0,width=\textwidth]{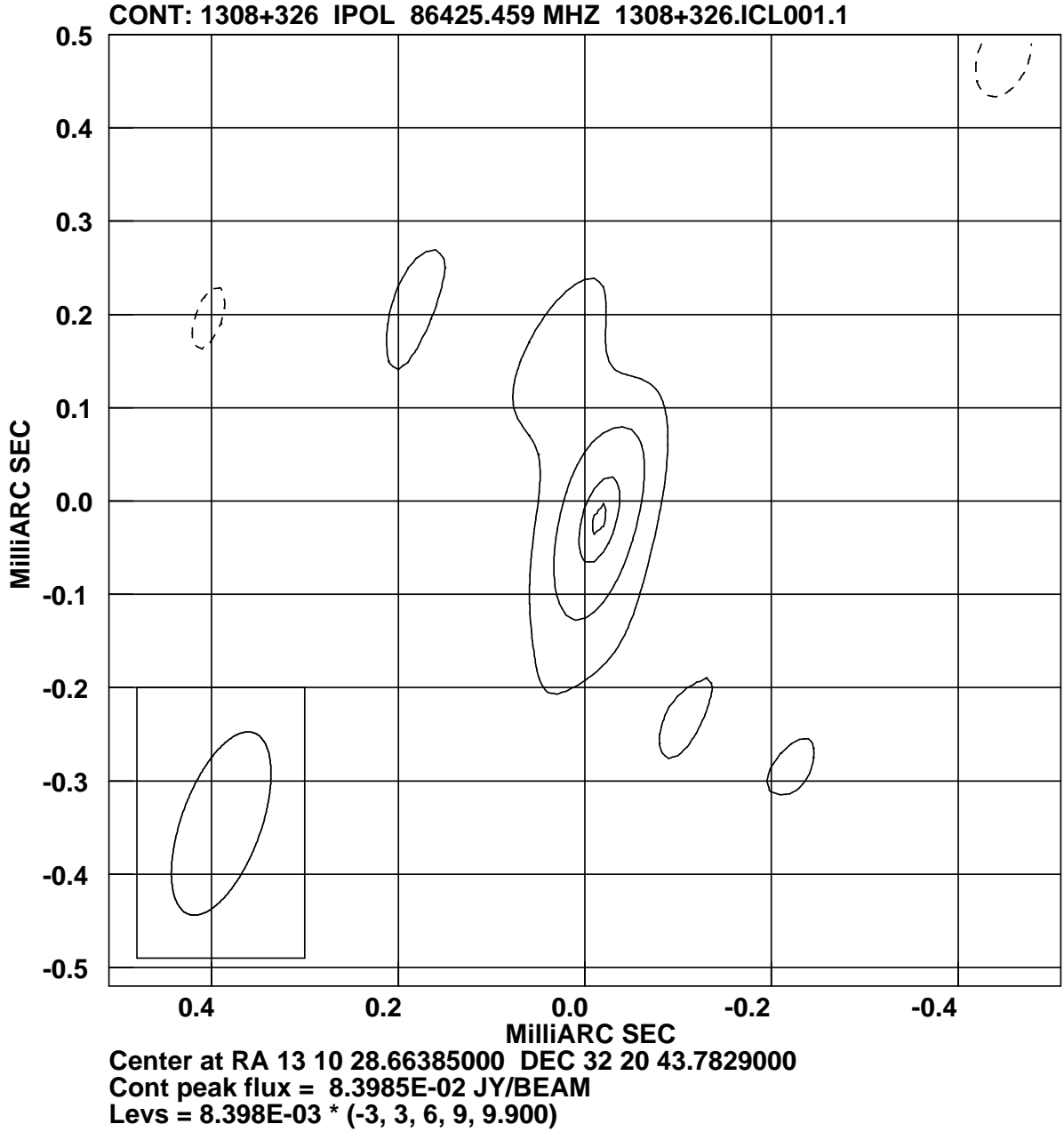}
\caption{Source-Frequency Phase Referenced ({\sc sfpr}-ed) map of
  1308+326 at 86 GHz from BD119, from VLBA observations at 43 and 86 GHz, along
  with a calibrator source, 1308+328, 14$^\prime$ away. The {\it rms}
  noise level of the map is 9 mJy/beam.  The flux recovery is 38\%. An
  astrometric offset of 22$\pm$13$\mu$as is measured, which is not
  significant.  Previous observations of this pair of source (see
  text) are compatible with a zero core-shift between 43 and 86 GHz,
  as seen in this image. We estimate an astrometrical error of $\sim
  20 \mu as$ for {\sc sfpr} observations with a close pair of sources.}
\label{1306_sfpr}
\end{figure}

Figure \ref{1306_sfpr} shows the {\sc sfpr}-image of 1308+326 at 86
GHz. This map was made using the calibration derived from the
BD119 observations of the same source at 43-GHz and further corrections
derived from observations of 1308+328, $14^\prime$ away, as described
above.  The peak flux in the {\sc sfpr-}image is 84 mJy, which
corresponds to a $\sim 38\%$ flux recovery, equivalent to a {\it rms}
phase error of $\sim \,$1.4 radians.  
The observational error estimate is in good quantitative agreement
with the theoretical predictions from our error analysis, as listed in
Table \ref{tab:two}, which arise from residual short term tropospheric
temporal fluctuations (i.e. {\it dynamic} tropospheric errors) and a
frequency switching cycle of $\sim 1$ minute.
The effect of these errors is expected to produce degraded images,
with the image signal-to-noise ratio reduced by the random phase
noise. For comparison, the only conventional PR map which has been
done at 86-GHz \citep{porcas_02_pr86}, using this same pair of
sources, resulted in a flux recovery of only 20\%.  However, the peak
of brightness in the map would not be shifted, hence the astrometric
errors are expected to be small, under 30$\,\mu$as, for the given
phase noise and assuming a synthesized beam $\sim\, 150 \, \mu$as.  We
measured the offset of the peak of brightness in the {\sc sfpr-}map
with the AIPS task {\sc jmfit} to be 22$\,\mu$as along a PA=-142$^o$,
with a formal fitting error of 13$\,\mu$as.
Therefore we conclude that there is no significant astrometric offset
in our {\sc sfpr}-map.  The results from previous astrometric
observations of this pair of sources
\citep{rioja_96_1308a,rioja_96_1308b,porcas_00_1308} at lower
frequencies are compatible with a zero core-shift between 43 and 86
GHz, as found in our analysis.

The multiple 1-hour long observations (BD123), carried
out on different days, of the 1308+326 and 1308+328 pair allowed us to
check the repeatability of the astrometric results and characterize
the robustness of the method with respect to weather.  We found that
in poor weather conditions the flux recovery fell dramatically, as
expected, since the frequency switching time was insufficiently fast
to follow the tropospheric fluctuations. However, in all cases most
baselines could be calibrated, and the positions of the peak of
brightness in the {\sc sfpr-}maps at 86 GHz were distributed around
the center of the map, with an average position and {\it rms} equal to
$20\pm17\,\mu$as.  Based on this repeatability we conclude that the
astrometric measurements using SFPR techniques are robust in various
weather conditions.  Additionally, this repeatability serves to give a
realistic error estimate of the astrometric precision achieved at 86
GHz using SFPR techniques, for a close pair of radiosources, of the order of
$\sim 20\, \mu$as.

\begin{figure}
\includegraphics[angle=0,width=\textwidth]{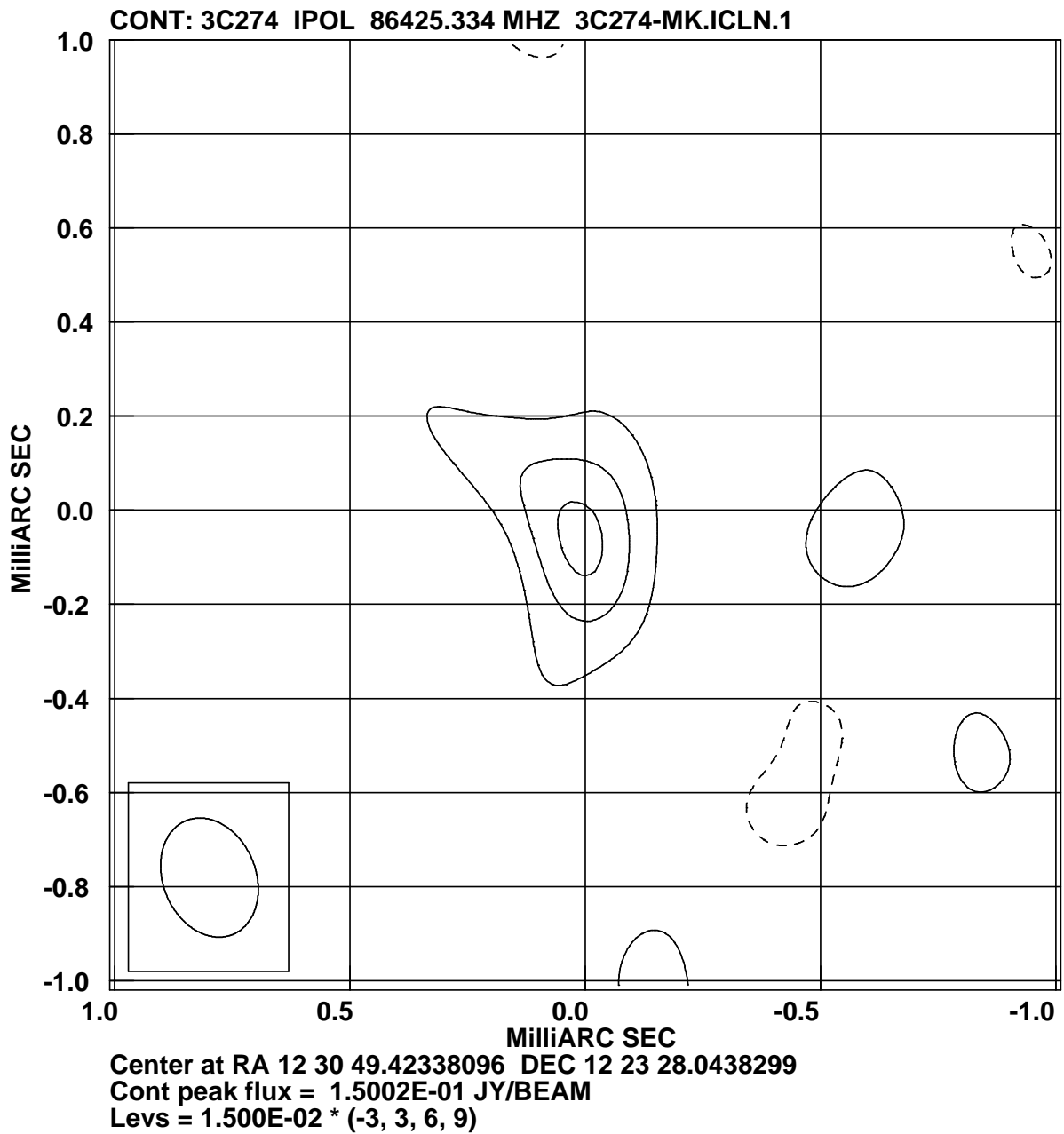}
\caption{Source-Frequency Phase Referenced ({\sc sfpr}-ed) map of
  3C274 at 86 GHz from BD119, from VLBA observations at 43 and 86 GHz, along with
  a calibrator source, 3C273, 10$^o$ away. The data from MK antenna
  have been edited out. The {\it rms} noise level of the map is 18 mJy/beam.  The
  flux recovery is 32\%. The direction of the astrometric offset
  (160$^o$) of the peak-flux with respect to the center of the map is
  compatible with that expected from the core shift, and its magnitude
  (48$\pm$30$\mu$as) is consistent with theoretical predictions for
  these sources. However it also agrees with expected propagation of
  astrometric errors. In the absence of other observations to compare
  our results with, we propose to use the offset of the peak of
  brightness with respect of the center of the {\sc sfpr-}map as a
  conservative upper limit for the astrometric accuracy achieved with
  SFPR techniques applied to pairs of sources with wide angular
  separation.}
\label{3c_sfpr}
\end{figure}

Figure \ref{3c_sfpr} shows the {\sc sfpr}-image of 3C274 at 86\,GHz.
This map was produced using the calibration derived from the
BD119 observations of the same source at 43-GHz and further corrections
derived from observations of 3C273, $10^o$ away.  The larger angular
separation between the sources in the 3C pair, compared to the
$14^\prime$ for the 1308+32 pair, along with the lower declination
and the North-South relative orientation,
makes this case the ultimate limit of what we would consider to be a
suitable {\sc sfpr} calibrator.  We note that if the goal was to make
accurate phase referenced maps, rather than test the calibration
scheme, M84 ($\sim 1.4^o$ away from 3C274) would be a more suitable
calibrator source.
The peak flux in the {\sc sfpr-}image is 150 mJy, which corresponds to a 
$\sim 32\%$ flux recovery compared to the peak in the hybrid map, which is
equivalent to a {\it rms} phase error of $\sim \,$1.5 radians.
This is similar to the flux recovery for the close pair, and in
agreement with the theoretical estimates from our error analysis
(Table \ref{tab:two}), since the dominant errors (i.e. $\sigma
\phi_{{\rm dtrp}}$) are only slightly dependent on the source separation and
both pairs where observed under identical weather conditions.  On the
other hand, the larger separation between the two sources of this pair
results in a significant residual ``{\it static}'' ionospheric error
(see Table \ref{tab:two}), which is expected to propagate into astrometric errors.
Also we find differences in the astrometric estimates which depend on the 
inclusion of data from the MK antenna or not, presumably since it
provides the longest baselines, and includes the lowest antenna
elevations.
Leaving out the MK data, the offset of the peak of brightness with
respect to the center in the {\sc sfpr-}map, measured with {\sc
  jmfit}, is 43$\,\mu$as along a PA=+160$^o$, with formal fitted
errors 30$\mu$as; this shift was doubled in magnitude when we
include MK data in the analysis. 

In the absence of any other astrometric observations at 86 GHz to
compare our results against, we consider here three scenarios to
interpret the measurements in the {\sc sfpr-}map.  Firstly, any
apparent source position shifts resulting from differential structure
blending effects from observations at different frequencies, or real
shifts due to opacity changes (true ``core-shifts''), are expected to
occur up stream along the jet structure axis. The source axis for
3C273 is along PA $\sim -142^o$ (Fig. \ref{3c_hybrid}), for 3C274 we
consider a range of PAs between $\sim -90^o/-45^o$ derived from our
maps (Fig. \ref{3c_hybrid}) and larger scale maps, respectively.
The {\sc sfpr} analysis measures the relative position shift in 3C274
between 86 and 43 GHz, with respect to that in 3C273.
Hence, given the orientation of the expected shifts at each source,
the combined shift is expected to appear 
in a range of PAs between +90$^o$/+135$^o$ (if dominated by 3C274) and
-142$^o$ (if dominated by 3C273), depending on the magnitudes of the
individual contributions. The direction of the measured shift in the
{\sc sfpr-}map (Fig. \ref{3c_sfpr}), along PA $\sim -164^o$, falls
in this range.
Secondly, we use the theoretical predictions for the core-shifts
between 43 and 86 GHz for these sources to estimate the corresponding
offset in the {\sc sfpr-}map. These are 65\,$\mu$as \citep{lob_98_cj}
for 3C273, and zero for 3C274 (Lobanov, personal comms.), which
correspond to an offset in the {\sc sfpr}-map of $\sim 65\,
\mu$as along PA $\sim -142^o$.
This theoretical position is $\sim 55\,\mu$as away (approximately 2$\sigma$,
based on the formal {\sc jmfit} errors) from that measured in the
SFPR-map shown in Figure \ref{3c_sfpr}.
Finally, we consider the propagation of astrometric errors in the analysis.
The residual systematic long-term ionospheric errors, as
listed in Table \ref{tab:two}, are expected to propagate into the astrometry.
The simulation studies carried out by \citet{pradel_06} show that the
dominant propagation of astrometric errors, in PR observations, are
along the declination coordinate for pairs of sources with low declination and
North-South relative orientation, as it is the case for 3C273/3C274.
Leaving aside the differences between the observational methods, and
extrapolating their findings to {\sc sfpr}, all or part of the
measured offset in the {\sc sfpr-}map could be due to this effect.

In conclusion, as we are unable to compare our findings with other
experimental results from any other technique, we propose to use the
observed shift in the {\sc sfpr-}map as an upper bound for the
astrometric uncertainties in the analysis method, when the angular
separations are of many degrees.  That is, we give a conservative upper
bound of 0.1 mas to the astrometric {accuracy} at 86 GHz using SFPR
techniques, in the case of two sources with a separation of $\sim
10^o$.

\section{Discussion}\label{sec:disc}

\subsection{\it Validation of SFPR method:} 

We have developed a new two-step calibration technique called SFPR
that, by precisely compensating for the effect of the propagation
medium in VLBI observations, enables high precision astrometry even at
the highest frequencies where conventional PR techniques fail.
Previous attempts using fast frequency-switching observations achieved
an increased coherence time as a result of the dual-frequency
tropospheric calibration, which enabled the detection of weak sources
but failed to provide astrometry due to remaining dispersive errors
\citep{middelberg_05_fs}.
Our method addresses this issue with a second 
calibration step, that corrects for those and hence enables
astrometry.  We have provided experimental demonstration of the
ability of the SFPR method to disentangle the astrometric signature
from the other contributions using VLBA observations at 86 GHz, the
highest VLBA frequency, and 43 GHz. This ``chromatic'' astrometric
signature corresponds to the
angular separation between the emitting regions at the two observing
frequency bands in the target source, assuming an achromatic
calibrator.  Therefore, our SFPR method is a valuable tool for studies
that require comparison and {\it bona-fide} astrometric registration
of images at two or more frequencies, even at the highest frequencies
possible with VLBI.  Additionally, when combined with conventional PR
observations at $\nu^{{\rm low}}$, SFPR can provide 
`PR-like' astrometry at $\nu^{{\rm high}}$.  That is, measurements 
of the positions
with respect to an external calibrator.  Such astrometric measurements
can be used for position stability, proper motion and parallax studies
at frequencies beyond the traditional limit of phase referencing
($\sim$43 GHz).  In summary this method offers the means to expand the
benefits that conventional PR techniques offer in the moderate
frequency regime, into the highest frequencies used in VLBI.  In
previous works dual frequency observations
have been used for the detection of weak sources at mm-VLBI; 
now, with the SFPR technique, {\it bona-fide} high precision astrometric 
mm-VLBI (and sub-mm) can also be performed. 

We have carried out an analytical error analysis to characterize the
performance of the SFPR method. This analysis shows that when using
frequency-switching observations the dominant source of errors are the
random fluctuations in the {\it dynamic} component of the
troposphere. Also that the magnitude of this error depends on the
frequency switching cycle at the observations and the weather
conditions, irrespective of the pair angular separation.  These
findings are in good qualitative and quantitative agreement with the
experimental results from our SFPR observations of two pairs of
sources with very different angular separations. This validates the
new astrometric method and the error analysis, confirming its
potential in mm-VLBI.  Furthermore it gives confidence in our
extrapolations into domains yet untested, that is to the sub-mm
regime, and to simultaneous dual frequency observations.

\subsection{\it Broad scope of application:}

The constraints for successful SFPR observations are relatively easy
to fulfill. The SFPR method has been successfully demonstrated with
VLBA observations of a pair of sources with a large (10$^o$) angular
separation, at 43/86\,GHz.  These results are encouraging as they
suggest that the SFPR method would work with any other combination of
integer-ratio frequencies provided suitable frequency switching cycles
are used, and that the angular separation between the calibrator and
target sources can be large and telescope switching cycles long;
certainly much more than those required for conventional PR.
The key elements for success of the new method in the high frequency
regime are, firstly, that the frequency switching operation in SFPR
can be carried out much faster than the source switching counterpart
in PR. Secondly, that finding a suitable SFPR calibrator source is
relatively easy, unlike for PR, because wider angular
separations and longer switching cycles are allowable, along with the extended
coherence at the high frequencies.  The SFPR
method can be implemented as a regular observing mode 
with existing instruments,
such as the VLBA, which support fast frequency switching
operations. The best astrometric performance is achieved when multiple
frequency bands can be observed simultaneously, as this provides an
exact tropospheric correction in all weather conditions, eliminates
the need of phase connection and increases the on-source time.  The
Korean VLBI Network (KVN), equipped with multi-channel receivers at
22/43/86/129 GHz, and telescopes like Yebes and Haystack are among the
instruments that can carry out simultaneous observations of multiple
high frequency bands to achieve maximum benefit from the
SFPR technique.
Based on our analysis it is clear that simultaneous frequency
observations can be very useful in high frequency VLBI. We strongly
suggest that this capability is included in all next-generation
instruments.
We conclude that this method is broadly applicable to mm-VLBI
observations of many target sources, and unique in providing {\it
  bona-fide} astrometrically registered images and high precision
relative astrometric measurements using existing and newly built
instruments.

\subsection{\it Applications to space-VLBI:}

Additionally to the errors in the atmospheric propagation models, the
geometric errors also introduce inaccuracies in the PR analysis, which
can ultimately prevent its application.  While for VLBI ground arrays
the telescope coordinates can be accurately measured with dedicated
geodesy campaigns this is of particular concern for space-VLBI
observations, since the precise orbit determination for a satellite
antenna is much more complicated.  For example, for VSOP-2, the
accuracy in the orbit reconstruction required for successful PR
observations at 43\,GHz must be better than 10-cm (A07), which is
challenging. For comparison, the typical orbit determination accuracy
for its predecessor, the HALCA satellite, was 2--5 meters using
Doppler measurements from the Ku-band link
\citep{porcas_vsop_00,rioja_vsop_09}.  Strategies to achieve the 10-cm
level of accuracy using global satellite navigation systems and
satellite laser ranging techniques are presented in
\cite{asaki_orbit}.

Alternatively, the SFPR method automatically 
corrects for any geometric errors, 
including any orbit determination errors, irrespective of their magnitude.
There are no specific requirements on the orbit accuracy for SFPR
analysis other than those imposed by the correlator fringe field of view,
which is typically many meters.
In addition to the astrometric applications, the increased
sensitivity resulting from longer coherence time is very
useful because of the limited size of an orbiting antenna, particularly
at the higher frequencies.
Also, the fast frequency-switching operation for SFPR is less demanding
than fast source-switching for PR, reducing the requirements on the
satellite attitude control system. Further discussions can be found in
\citet{memo_32}.  Therefore, we believe this method will be very
useful for space VLBI missions. In particular, applied to VSOP-2
observations, it would enable increased sensitivity allowing the
detection of weaker sources, and permit astrometric measurements
and long term monitoring projects at 43 GHz,
by using the calibration derived from interleaving observations at 22 GHz
even with a coarse orbit determination. For other future
space VLBI missions at high frequencies, e.g. ``Millimetron'' 
\citep{millimetron}, 
having simultaneous observations at multiple frequencies 
combined with all or some aspects of the SFPR calibration techniques 
would enable enhanced sensitivity and astrometric capabilities.

\subsection{\it Astrometric Precision \& Applications} 

High precision astrometric and sensitive measurements are valuable
tools to provide insight into astrophysical phenomena, as demonstrated
by application of phase referencing techniques at a moderate frequency
regime (i.e. up to 43 GHz).  The SFPR method enables such measurements
in the high frequency regime by means of an improved atmospheric
calibration.  Our analytical error analysis shows that the
tropospheric {\it static} component is readily compensated using SFPR
techniques and that, in general the {\it dynamic} component would be
the dominant source of errors with fast frequency switching
observations, leading to $\sim\,$a few tens of micro-arcseconds
astrometric precision. Based on the repeatability of results from
observations with the VLBA we estimate an astrometric precision of
$\sim \, 20\,\mu$as.  With simultaneous dual frequency observations
both the {\it static} and {\it dynamic} components of the troposphere
would be precisely compensated, and the much smaller ionospheric
residuals become the dominant source of errors.  Increasing the
reference frequency can result in negligible ionospheric residuals,
therefore SFPR techniques at high frequencies offer the prospect of
achieving the theoretical astrometric precision set by the
interferometer beam size and the signal-to-noise ratio
\citep{thompson_isra_2}.

SFPR techniques applied to VLBI spectral line maser observations
would allow a precise {\it bona-fide} astrometric spatial registration
of, for example, the SiO emission structures at different
frequency bands (43, 86, 129 GHz) in the same source for studies of
the circumstellar environment in AGB stars.
When applied to AGN studies, mm and sub-mm VLBI observations probe the
inner-jet regions. SFPR can add the measurement of
core-shifts with micro-arcsecond precision, plus enable deeper
observations for the detection of weak sources. 
For astrometric measurements relative to an external reference, in
combination with conventional PR at $\nu^{{\rm low}}$, the final precision
would be the quadratic sum of errors from both techniques.  This would
allow the application to proper motion studies of maser emission at the
high frequencies, and precise astrometric monitoring programs of the `jet
foot-prints' to unveil the cause of the observed jet-wobbling
phenomena.

\section*{Acknowledgments}

RD acknowledge support for this research by a Marie Curie
International Incoming Fellowship within the EU FP6 under contract
number MIF1-CT-2005-021873, undertaken at the OAN.
VLBA is operated by the National Radio Astronomy Observatory, which is
a facility of the National Science Foundation operated under
cooperative agreement by Associated Universities, Inc.


\clearpage
\begin{landscape}
\begin{table}
\caption{Estimated residual phase error budget for
  Source-Frequency Phase Referencing (SFPR) techniques, per baseline,
  using the formulae in Section 3.  For comparison
  we include the error budged estimates for Phase Referencing (PR)
  techniques at 43 GHz, using formulae in A07.  For SFPR the errors
  correspond to those at the higher frequency.  In all cases weather
  conditions were 
  set to `good' and the tropospheric zenith delay error $\Delta l_z = 3\,$cm
  (except when marked with ($^b$), then 1 cm). All other parameters are
  set to the nominal values in the equations.}\label{tab:one}
\begin{tabular}{|ll|c|c||c|c||c|}
\hline
  & & \multicolumn{5}{c|}{\bf RMS phase [deg]}  \\
  \cline{3-7} \multicolumn{2}{|c|}{Error term} &
  \multicolumn{2}{c||}{SFPR$_{43\rightarrow86 {\rm GHz}}$}
  &  \multicolumn{2}{c||}{SFPR$_{43\rightarrow129 {\rm GHz}}$}& {PR$_{{\rm 43 GHz}}$}  \\
       &   & {\sc Source/Freq.} & {\sc Only Source} & {\sc Source/Freq.} & {\sc Only Source}&  {\sc Source}\\
       &   & {\sc switching}  & {\sc switching}  & {\sc switching}  & {\sc switching}  & {\sc switching}\\
 & & (1) & (2) & (3) & (4) & (5) \\
\hline
Dynamic Troposphere & $\sigma \phi^{{\rm high}}_{{\rm dtrp}}$    & 76     & 0 & 115 & 0 & 43        \\
Static Troposphere & $\sigma \phi^{{\rm high}}_{{\rm strp}}$    & 0      & 0 & 0   & 0& 107 (36$^b$) \\ 
Dynamic Ionosphere & $\sigma \phi^{{\rm high}}_{{\rm dion}}$    & 1
(2$^a$)    & 1   (2$^a$) & 2  (3$^a$)   & 2  (3$^a$) & 0.3         \\
Static Ionosphere & $\sigma \phi^{{\rm high}}_{{\rm sion}}$    & 6 (29$^a$) & 6 (29$^a$) & 10 (51$^a$)  & 10 (51$^a$)& 4         \\
Geometric$^c$ & $\sigma \phi^{{\rm high}}_{{\rm geo}}$     & 0      & 0 & 0   & 0 & 30        \\
Thermal Noise$^d$ & $\sigma \phi^{{\rm high}}_{{\rm thermal}}$  & 1.5 & 1.5 & 2.6 & 2.6& 0.2  \\
\hline
{\sc sum}  &            & 77 (82$^a$)& 6 (29$^a$)& 115 (125$^a$)& 10 (51$^a$)& 120 (63$^b$) \\
\hline 
\end{tabular}

\vspace*{1cm}

{\footnotesize {\bf (1):} {SFPR observations at 43 and 86 GHz,
    using {\underline {\it frequency and source switching}}, with
    frequency switching cycle ${\rm T^{\nu}_{swt}} = 60\,$seconds, and
    source switching cycle ${\rm T_{swt}}= 300\,$seconds.  The source
    switching angles are $\Delta \theta = 2^o$ and $\Delta \theta =
    10^o$, with values for the later in brackets and labelled with ($^a$)
    if different from those for the former.}
   {\bf (2):} {SFPR observations at 43 and 86 GHz, using
     simultaneous dual frequency observations and {\underline {\it only
         source switching}}, with a (source switching) cycle ${\rm
       T_{swt}}= 300\,$seconds. The source switching angles are $\Delta \theta
     = 2^o$ and $\Delta \theta = 10^o$, with values for the later
     in brackets and labelled with ($^a$) if different from those for the former.}  
   {\bf (3):} {Same as {\bf (1)}, for 43 and 129 GHz.} 
   {\bf (4):} {Same as {\bf (2)}, for 43 and 129 GHz.} 
   {\bf (5):} {PR observations at 43 GHz, with a switching angle
     $\Delta \theta = 2^o$ an source switching cycle ${\rm T_{swt}}=
     60\,$seconds.}  
   {\bf (a):} {Error term for $\Delta \theta = 10^o$, given
     only if different from that for $\Delta \theta = 2^o$, for SFPR.}  
   {\bf (b):} {PR with improved tropospheric calibration 
   strategy to achieve $\Delta l_z$=1cm.} 
   {\bf (c):} {The geometric errors $\sigma \phi_{{\rm geo}}$ 
      are a combination of the $\sigma \phi_{{\rm bl}}$ and $\sigma
      \phi_{\Delta s}$ errors with nominal values.}
    {\bf (d):} {The $\sigma \phi_{{\rm thermal}}$ contribution
      has been calculated using the SEFD parameter values for the VLBA
      at 43 and 86 GHz, and for the KVN at 129 GHz, and source fluxes
      of 0.1\,Jy.}}

\end{table}\end{landscape}

\clearpage

\begin{table}
\caption{Estimated residual phase error budget for Source-Frequency
  Phase Referencing (SFPR) techniques, per baseline, for our
  observations of the two pairs of sources, at 43/86 GHz (matching BD119). Also, for comparison, 
  estimated errors for PR observations at 86 GHz for the close pair (PR is not feasible for the pair $10^o$ apart). 
  The error values have been calculated using the
  formulas in this paper, for SFPR, and in A07, for PR. For SFPR the
  errors correspond to those at the higher frequency. In all cases
  weather conditions were set to `good' and the tropospheric zenith
  delay error $\Delta l_z = 3\,$cm.}\label{tab:two}
\begin{tabular}{|ll|c|c||c|}\hline
 & & \multicolumn{3}{c|}{\bf RMS phase [deg]}  \\ 
\cline{3-5} 
 & & \multicolumn{2}{c||}{1308+326/1308+328}  &  3C273/3C274 \\
& & \multicolumn{2}{c||}{$\delta \sim 32^o$, 14$^\prime$ apart} &  $\delta \sim 0^o$, $10^o$ apart \\
\multicolumn{2}{|c|} {Error term} & {\sc SFPR} & {\sc Conventional PR} & {\sc SFPR} \\
  &         & {$43\rightarrow 86 {\rm GHz}$}  & {\rm 86 GHz}  &
           {$43\rightarrow 86 {\rm GHz}$}  \\
  & & (1) & (2) & (3) \\
\hline
Dynamic Troposphere & $\sigma \phi^{{\rm high}}_{{\rm dtrp}}$    & 68     & 69 & 80    \\
Static Troposphere & $\sigma \phi^{{\rm high}}_{{\rm strp}}$    & 0      & 12 & 0      \\ 
Dynamic Ionosphere & $\sigma \phi^{{\rm high}}_{{\rm dion}}$    & 0.3    & 0.1 & 3         \\
Static Ionosphere & $\sigma \phi^{{\rm high}}_{{\rm sion}}$    & 0.3    & 0.1  & 37        \\
Geometric$^a$ & $\sigma \phi^{{\rm high}}_{{\rm geo}}$     & 0      & 3  & 0       \\
Thermal Noise$^b$ & $\sigma \phi^{{\rm high}}_{{\rm thermal}}$ & 0.7 & 0.5 & 0.2   \\
\hline
{\sc SUM}   &         & 68 & 70 & 88   \\
\hline

\end{tabular}

\vspace*{1cm}

{\footnotesize {\bf (1):} {SFPR observations at 43/86 GHz 
    of the pair of sources
    1308+326 and 1308+328 using a frequency switching cycle 
    ${\rm T^{\nu}_{swt}} = 60\,$seconds,
    and source switching cycle ${\rm T_{swt}}= 300\,$seconds. 
    The switching angle is  $\Delta \theta = 14^\prime$.
    Values of parameters $Z_g, Z_i, Z_f$ are $30^o$, $29^o$ and
    $28^o$, respectively.}
{\bf (2):} {Conventional PR observations at 86 GHz 
    of the pair of sources 1308+326 and 1308+328, 14$^\prime$ away, 
    with a switching cycle ${\rm T_{swt}}= 60\,$seconds.} 
{\bf (3):} {SFPR observations at 43/86 GHz 
    of the pair of sources 3C273/3C274 using a frequency switching cycle 
    ${\rm T^{\nu}_{swt}} = 60\,$seconds,
    and source switching cycle ${\rm T_{swt}}= 300\,$seconds. 
    The switching angle is  $\Delta \theta = 10^o$.
    Values of parameters $Z_g, Z_i, Z_f$ are $50^o$, $47^o$ and
    $46^o$, respectively.}
{\bf (a):} {This contribution has been calculated using $\sigma \phi_{{\rm bl}}=0.5$cm.}
{\bf (b):} {The $\sigma \phi_{{\rm thermal}}$ contribution
      has been calculated using the SEFD parameter values for the VLBA
      at 43 and 86 GHz (1500 and 4000 Jy, respectively), and the source
      fluxes are the peak flux measured in the corresponding hybrid maps
      shown in Figures \ref{1308_hybrid} and \ref{3c_hybrid}.}}
\end{table}
\end{document}